\documentclass[10pt, twocolumn]{article}

\usepackage[a4paper, inner=1.5cm, top=1.5cm, outer=1.5cm, bottom=1.5cm, includefoot]{geometry}
\usepackage{amsmath}
\usepackage{bm}
\usepackage[max2]{authblk}
\usepackage{amssymb}
\usepackage{mathrsfs}
\usepackage{graphicx}
\usepackage{enumitem}
\usepackage{tikz-network} 
\usepackage[ruled,lined,algosection]{algorithm2e}
\usepackage[style=ieee,natbib=true,url=false,isbn=false,mincitenames=1,maxcitenames=2]{biblatex}
\usepackage{amsthm}
\usepackage{appendix}
\usepackage{subdepth}
\usepackage{float}
\usepackage{hyperref}
\usepackage{mathtools}
\usepackage{multirow}
\usepackage{array}
\usepackage{mathtools}
\usepackage{diagbox}
\usepackage{placeins}
\usepackage{abstract}
\usepackage{titling}

\DeclareUnicodeCharacter{03B3}{$\gamma$}
\DeclareUnicodeCharacter{2212}{--}
\DeclareUnicodeCharacter{03B1}{$\alpha$}

\newcommand{\ubar}[1]{\mkern 0.5mu\underline{\mkern-0.5mu#1\mkern-0.5mu}\mkern 0.5mu}
\newcommand{\uubar}[1]{\ubar{\ubar{#1}}}

\let\epsilon\varepsilon
\let\mtheta\theta
\let\theta\vartheta
\let\mrho\rho
\let\rho\varrho
\let\mphi\phi
\let\phi\varphi
\let\Gamma\varGamma
\let\Delta\varDelta
\let\Theta\varTheta
\let\Lambda\varLambda
\let\Xi\varXi
\let\Pi\varPi
\let\Sigma\varSigma
\let\Upsilon\varUpsilon
\let\Phi\varPhi
\let\Psi\varPsi

\let\Omega\varOmega

\newcommand{\grad}[3]{\nabla_{#1}^{#2} #3} 

\newcommand{\clE}{\mathcal{E}}

\newcommand{\sT}{\mathsf{T}} 

\newcommand{\scI}{\mathscr{I}}

\newcommand{\scL}{\mathscr{L}}

\newcommand{\scP}{\mathscr{P}}

\newcommand{\uf}{\ubar{f}}
\newcommand{\ug}{\ubar{g}}

\newcommand{\ur}{\ubar{r}}

\newcommand{\uu}{\ubar{u}}
\newcommand{\uv}{\ubar{v}}

\newcommand{\uuA}{\uubar{A}}

\newcommand{\uuS}{\uubar{S}}
\newcommand{\uuT}{\uubar{T}}

\newcommand{\uusigma}{\uubar{\sigma}}

\newcommand{\<}{\langle}
\renewcommand{\>}{\rangle}

\newcommand{\uuubar}[1]{\ubar{\ubar{\ubar{#1}}}}
\newcommand{\uuuT}{\uuubar{T}}

\newtheorem{remark}{Remark}

\newcolumntype{x}[1]{>{\centering\arraybackslash\hspace{0pt}}p{#1}}

\newcommand{\Exp}[1]{\left\langle #1 \right\rangle}

\newtheorem{example}{Example}

\title{Exact average many-body interatomic interaction model for random alloys}
\author{M. Hodapp\thanks{maxludwig.hodapp@mcl.at}}
\affil{Materials Center Leoben Forschung GmbH (MCL), Leoben (AT)}

\renewcommand\footnotemark{}

\begin{document}

\twocolumn[

\begin{@twocolumnfalse}

\maketitle

\begin{abstract}
    Understanding the physical origin of mechanisms in random alloys that lead to the formation of microstructures requires an understanding of their average behavior and, equally important, the role of local fluctuations around the average.
    Material properties of random alloys can be computed using direct simulations on random configurations.
    However, some properties are very difficult to compute, for others it is not even fully understood how to compute them using random sampling, in particular, interaction energies between multiple defects.
    To that end, we develop an atomistic model that does the averaging on the level of interatomic potentials.
    With such an average interatomic interaction model the problem of averaging via random sampling is bypassed since the problem of computing material properties on random configurations reduces to the problem of computing material properties on single crystals, the \emph{average alloy}, which can be done using standard techniques.

    To be predictive, we develop our average model on the class of linear machine-learning interatomic potentials (MLIPs).
    To that end, using tools from higher-order statistics, we derive an analytic expansion of average many-body per-atom energies of linear MLIPs in terms of average tensor products of the feature vectors of the underlying machine-learning model that scales linearly with the size of an atomic neighborhood.
    In order to avoid forming higher-order tensors composed of products of feature vectors, we develop an implementation using equivariant tensor network potentials, a class of linear MLIPs, that contracts the feature vectors to small-sized tensors, and then takes the average.
    We validate our average potential by demonstrating the convergence of direct Monte Carlo simulations to the exact value for properties of the NbMoTaW medium-entropy alloy.
    Moreover, we show that it predicts the compact screw dislocation core structure, in agreement with density functional theory, as opposed to state-of-the-art average embedded atom method potentials that predict artificial polarized cores.
    Hence, we anticipate that our model will become useful for understanding mechanistic origins of material properties and for developing predictive models of mechanical properties of random alloys.
\end{abstract}

\end{@twocolumnfalse}

]

\saythanks

Random alloys are a broad class of materials that does neither impose any restrictions on the number of atomic species nor its composition \cite{miracle_critical_2017}.
For example, high-entropy alloys, arguably the most prominent subclass of random alloys, are usually composed of five or more principal species at close-to equiatomic composition.
Unlike intermetallics, random alloys do not form an ordered phase but a solid solution with some ordering, or perhaps no ordering at all.
The key difference between random alloys and conventional classes of alloys, such as steels, is the local disorder inherent to random alloys.
It is due to this local disorder that material properties of random alloys can strongly deviate locally from their average, potentially changing the occurrence of mechanisms \cite{george_high_2020}.
Hence, a fully predictive theory of random alloys must take \emph{average properties as well as deviations from them} into account.

One example where local disorder is expected to be relevant is fracture toughness.
Here, models like Rice-Thomson that are based on average properties tend to underpredict ductility trends \cite{mak_ductility_2021,singh_ductility_2023} and a potential reason for this behavior could be local fluctuations along the crack front that facilitate dislocation emission with respect to the average behavior \cite{singh_ductility_2023}.
However, there exists presently no model that describes this phenomenon from first principles.

\begin{figure*}[hbt]
    \centering
    \includegraphics[width=0.9\textwidth]{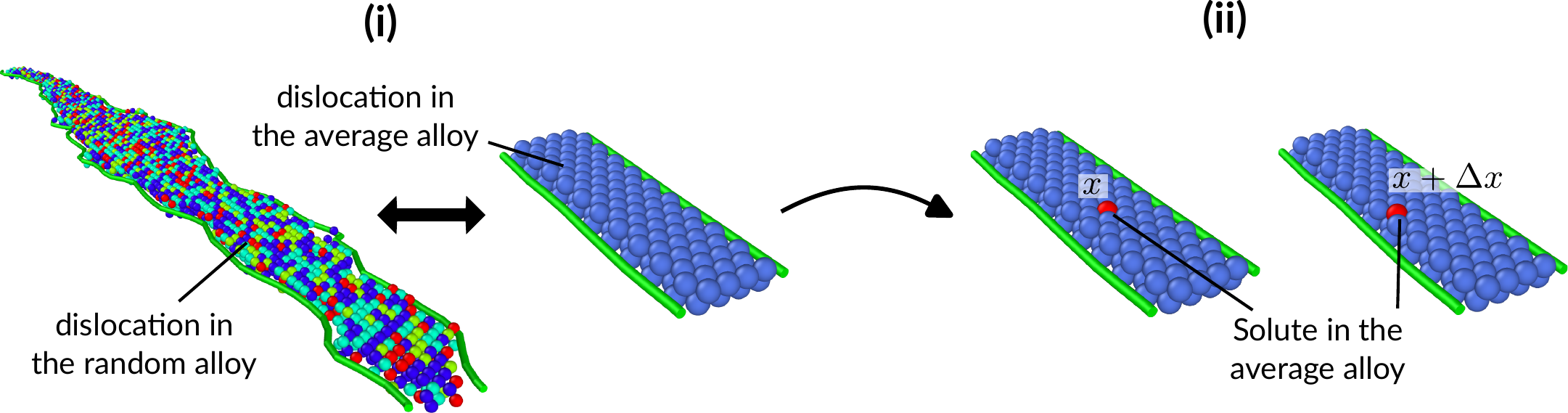}
    \caption{Schematic illustration of a materials modeling workflow using the average alloy concept.
    (i) Comparison of the dislocation motion in the random alloy and in the average alloy with respect to some target property, e.g., the Peierls stress.
    (ii) If the results differ, add real random alloy "solute" atoms into the average alloy at position $x$ and compute the energy change when moving the dislocation by $\Delta x$ relative to the solute.
    This energy change can then be transformed into the standard deviation that corrects the target property computed in the average alloy from (i).
    The figure of the dislocation in the random alloy is by courtesy of Wolfram N\"{o}hring}
    \label{fig:average_concept}
\end{figure*}

To that end, Curtin and co-workers introduced a conceptual framework that considers fluctuations in the energy landscape as deviations from some average energy landscape that represents an \emph{average alloy} (see, e.g., \cite{nohring_design_2020} for some review of this concept).
Within the average alloy framework (Figure \ref{fig:average_concept}), first outlined by \citet{varvenne_theory_2016}, a defect in the average alloy can interact with \emph{species composing the random alloy} that can be imagined as \emph{solutes in the average alloy}.
The problem of quantifying the influence of local disorder on the behavior of defects can thus be considered as a problem of the defects moving in a random field of solutes.
The energy of a defect $\Pi^{\rm d}$ can then be written as an additive split of the average energy and a sum over all defect-solute interaction energies $\Pi_i^{\rm d\text{-}s}$, which has zero average, such that
\[
    \Pi^{\rm d}
    =
    \Exp{\Pi^{\rm d}}
    +
    \sum_i \Pi_i^{\rm d\text{-}s}
    ,
\]
omitting the solute-solute interactions (which, in fact, have been shown to be negligible for refractory alloys \cite{nag_effect_2020}).
Very often, the quantity of interest is, however, not the total energy but the energy difference $\Delta\Pi^{\rm d}$ of moving the defect in a certain direction.
Under certain assumptions, the standard deviation of $\Delta\Pi^{\rm d}$ can be described solely in terms of the interactions $\Delta\Pi_i^{{\rm d}\text{-}X}$ of the original alloy species, denoted by $X$, with the defect inserted in the average alloy and thus reduces to the remarkably simple quantity \cite{varvenne_theory_2016}
\begin{equation}\label{eq:standard_deviations}
    \sigma
    =
    \left( \sum_i \sum_X c_X \left( \Delta\Pi_i^{{\rm d}\text{-}X} \right)^2 \right)^{1/2}
    ,
\end{equation}
where $c_X$ is the concentration of species $X$.

Up to now, this framework has been successfully applied to predict different mechanical and material properties of random alloys, such as solute strengthening in face-centered cubic and body-centered cubic high-entropy alloys \cite{varvenne_theory_2016,maresca_theory_2020}, cross-slip barriers in face-centered cubic solid solutions \cite{nohring_crossslip_2018}, or grain boundary roughening in medium-entropy alloys \cite{baruffi_theory_2022}.
The power of this framework lies in the simplicity of computing those deviations because interaction energies $\Pi_i^{\rm d\text{-}s}$ between defects and solutes are computed in the same way as they are computed for single crystals, which is well-established.

On the other hand, the major difficulty that comes along with starting a theory from an effective medium approach is finding a model that computes average properties and deviations from them.
In principle, one could bypass this step and instead use Monte Carlo sampling, but this procedure becomes very costly when considering extended defects like stacking faults (cf., \cite{varvenne_averageatom_2016}).
Crucially, there are quantities for which it is \emph{not even clear how to compute them} via random sampling, such as dislocation-solute interaction energies, or the average line tension of a dislocation.

An accurate method that allows to predict material properties of random alloys is density functional theory (DFT).
In particular, for disordered systems, an efficient DFT-based method that represents the effective medium is the coherent potential approximation (CPA) (e.g., \cite{ruban_configurational_2008}).
However, the CPA is an on-lattice method and can, therefore, not account for arbitrary relaxations.
Another popular effective medium method based on DFT is the virtual crystal approximation (e.g., \cite{faulkner_modern_1982}) that linearly interpolates between the pseudopotentials, weighted by the concentrations of the corresponding atomic species.
However, it is known that the virtual crystal approximation becomes unreliable when the pseudopotentials to be averaged are not strongly similar to each other \cite{faulkner_modern_1982}.
Hence, describing random alloys with DFT is thus far limited to a few properties.
For example, \citet{varvenne_theory_2016,rao_simpler_2024} derived approximations of the dislocation-solute interaction energies in terms of bulk properties, such as elastic constants, misfit volumes, etc., that can be computed within the CPA framework \cite{moitzi_accurate_2022}.

A general approach using interatomic potentials based on two-body interactions has been developed by \citet{smith_application_1989} in the context of the embedded atom method (EAM) \cite{daw_embeddedatom_1993}.
Two-body interactions can be written as linear combinations of pair-wise contributions
\begin{equation}\label{eq:pair-wise_term}
    \sum_j f(\ur^{ij}, s^i, s^j),
\end{equation}
where $\ur^{ij}$ is the difference between the positions of the $i$-th atom and its $j$-th neighbor, respectively, and $s^i$ and $s^j$ are the corresponding atomic species.
If $s^i$ and $s^j$ are uncorrelated, then $f$ can be averaged analytically leading to an \emph{average potential}.
While average EAM potentials have been shown to \emph{capture} many average properties of random alloys, and deviations from them, they are generally not able to accurately \emph{predict} material properties, such as stacking fault or surface energies (e.g., \cite{novikov_aiaccelerated_2022}).
Another well-known deficiency is the problem of average EAM potentials not being able to predict the compact screw dislocation core structures in several bcc random alloys, as predicted by DFT \cite{yin_initio_2020}, but rather show strongly polarized cores \cite{maresca_theory_2020}.
To develop more accurate potentials, the idea of averaging atomic interactions could, in principle, be applied analogously to terms with a higher body-order $d$,
\begin{equation}\label{eq:many-body_term}
    \sum_{j_1 \neq \ldots \neq j_d} f(\ur^{ij_1}, \ldots, \ur^{ij_d}, s^i, s^{j_1}, \ldots, s^{j_d}),
\end{equation}
that appear, e.g., in modified EAM potentials (e.g., \cite{lee_second_2001}), but proceeding this way quickly becomes infeasible because \eqref{eq:many-body_term} scales exponentially with the body-order.

An efficient way to incorporate many-body terms into interatomic potentials are machine-learning interatomic potentials (MLIPs) \cite{behler_generalized_2007,bartok_gaussian_2010,thompson_spectral_2015,shapeev_moment_2016,zhang_deep_2018,drautz_atomic_2019,pun_physically_2019,batzner_equivariant_2022,takamoto_teanet_2022,batatia_mace_2022} using pair-wise interactions \eqref{eq:pair-wise_term} as features.
This way, many-body terms in MLIPs are constructed using a separation of variables Ansatz as follows
\begin{equation}\label{eq:many-body_term_linear}
    \sum_{j_1} f_1(\ur^{ij_1}, s^i, s^{j_1}) \ldots \sum_{j_d} f_d(\ur^{ij_d}, s^i, s^{j_d}).
\end{equation}
With such a construction, the problem of computing many-body interactions reduces to the problem of computing products of sums, so, for a given body-order, computing the sum \eqref{eq:many-body_term_linear} scales linearly with the size of the neighborhood.
Moreover, it has been shown that MLIPs built from linear combinations of many-body terms like \eqref{eq:many-body_term_linear} are able to approximate local quantum-mechanical models down to the typical noise in numerical DFT codes (cf., \cite{shapeev_moment_2016}).
This favorable combination of efficiency and accuracy has recently triggered growing developments of new potentials for random alloys (e.g.,  \cite{li_complex_2020,byggmastar_modeling_2021,hodapp_machinelearning_2021,mccarthy_atomic_2023,cao_capturing_2024,song_generalpurpose_2024,moitzi_initio_2024}).
However, due to the self-interaction of pair-wise features appearing in \eqref{eq:many-body_term_linear}, they cannot be directly averaged analytically.

In this work, we develop a \textbf{framework for averaging linear MLIPs}.
Recognizing that linear MLIPs can be cast into multilinear forms, with the arguments being sums of pair-wise interaction terms, it follows that the problem of averaging linear MLIPs reduces to the problem of converting a product of sums into disjoint sums, which can be averaged as above for EAM potentials.
The averaged disjoint sums can then be converted back to products of sums leading to an \emph{average interatomic interaction model} that scales linear with the size of the atomic neighborhood.
However, in such an expansion, there appear terms that require computing average tensor products of feature vectors.
To avoid forming those higher-order tensors explicitly, we develop an implementation of the average model using equivariant tensor network (ETN) potentials \cite{hodapp_equivariant_2023}, a class of linear MLIPs that factorize the tensor of model coefficients into a product of low-order tensors, up to the order of three.
Using ETN potentials, the feature vectors can then be contracted with these low-order tensors before averaging, leading to smaller tensors whose size depends on the rank of the tensor network, which can be made much smaller than the size of the feature vectors without sacrificing accuracy.
We validate our implementation by demonstrating the convergence of Monte Carlo simulations to our exact values.
Moreover, we will show that our model does not suffer from the deficiencies inherent in average EAM potentials and is able to predict compact average dislocation core structures in random refractory alloys, in agreement with DFT.

\section*{Results}

\subsection*{Multilinear potentials}

We assume that the total energy $\Pi$ of an atomic configuration is a function of the atomic positions $\ur^i$ and the atomic species $s^i$.
The local neighborhood of the $i$-th atom is the set $\{ \{ \ur^{ij} \}, s^i, \{ s^j \} \}$ containing all relative positions $\ur^{ij} = \ur^j - \ur^i$ between the $i$-th atom and its $j$-th neighbor, the atomic species of the $i$-th atom, and the atomic species of all neighboring atoms.
We then assume that $\Pi$ can be partitioned into local per-atom contributions $\clE$ that depend on the neighborhood such that
\[
    \Pi
    =
    \sum_i
    \clE(\{ \{ \ur^{ij} \}, s^i, \{ s^j \} \})
    .
\]

In this work, we model the per-atom energies using polynomial potentials.
To represent polynomial potentials, we will use multilinear forms, and, therefore, we henceforth call them \emph{multilinear potentials}.
The reasoning behind this choice is that, as we shall see in the following section, multilinear forms are convenient representations of $\clE$ for expanding its average $\Exp{\clE}$ in terms of products of averages of tensor products of feature vectors.
A multilinear potential can be written as a contraction of a tensor product of $d$ feature vectors $\uv^k$, $\bigotimes_{k=1}^d \uv^k$, that depend on the atomic neighborhood, with an order-$d$ tensor $T_{\langle d \rangle}$ that contains the model coefficients and encodes the model symmetries (rotation and reflection), defined through the contraction operator $\times$ as follows
\begin{equation}\label{eq:multilinear_potential}
    \clE
    =
    T_{\< d \>}
    \times
    \left( \bigotimes_{k=1}^d \uv^k \right)
    .
\end{equation}
The dimension $d$ of $T_{\langle d \rangle}$ defines the body-order of the potential.
Multilinear potentials can provably approximate any regular potential energy landscape of a local quantum-mechanical model, provided that the feature vectors are complete (see the following example).
Examples from this class of potentials are, e.g., spectral neighbor analysis potentials (SNAPs) \cite{thompson_spectral_2015}, moment tensor potentials (MTPs) \cite{shapeev_moment_2016}, or atomic cluster expansion (ACE) potentials \cite{drautz_atomic_2019}.
For our implementation that we describe later on after presenting the averaging scheme, we will use equivariant tensor network (ETN) potentials \cite{hodapp_equivariant_2023}, a variant of multilinear potentials that appears to be particularly efficient for implementing contractions of higher-order tensors that arise in the expansion of average energies.

\begin{example}[Feature vector]
    We do not place any particular assumptions on the $v$'s but the perhaps most common way of defining a complete set of feature vectors is by separating radial and angular contributions as follows
    \begin{multline}\label{eq:example_feature_vector}
        v_k
        =
        v_{(n \ell m)} \\
        =
        \left\{
        \begin{aligned}
            1 & \quad n = \ell = m = 0, \\
            \sum_j Q_\alpha(\vert \ur^{ij} \vert) z_\beta^i z_\gamma^j Y_{\ell m}(\widehat\ur^{ij}) & \quad \text{else.}
        \end{aligned}
        \right.
        ,
    \end{multline}
    with the multi-index $n = (\alpha\beta\gamma)$, where the $Q_\alpha$'s are radial basis functions that smoothly go to zero beyond some cut-off radius of a few lattice spacings, $z_\beta^i,z_\gamma^j$ are vectors that define the species of the $i$-th and $j$-th atoms, respectively, and the $Y_{\ell m}$ are spherical harmonics;
    a way to define $z_\beta^i,z_\gamma^j$ is to assume a lexicographical order of atomic species $s = \{ 1, 2, 3, \ldots, N_{\rm spec} \}$ and define $z_\beta^i = \delta_{\beta s^i},z_\gamma^j = \delta_{\gamma s^j}$, where $\delta$ is the Kronecker delta.

    The feature vectors to be contracted may differ in the number and type of features, e.g., we may envision contracting positional features with magnetic moments;
    this is reflected in the functional form \eqref{eq:multilinear_potential}.
\end{example}

\begin{example}[Polynomial representation of $\clE$]
    The tensorial representation \eqref{eq:multilinear_potential} can always be converted into a multivariate polynomial.
    For illustration purposes, assume that $d = 2$ and that $\uv^1 = \uv^2 = \uv$, with
    $
        \uv = \begin{pmatrix} 1 & v_1 & v_2 \end{pmatrix}^\sT
    $.
    Then,
    \begin{multline*}
        \clE
        =
        \uv^\sT \cdot \left( \uuT \uv \right)
        =
        \begin{pmatrix} 1 & v_1 & v_2 \end{pmatrix}
        \begin{pmatrix} T_{11} & T_{12} & T_{13} \\ & T_{22} & T_{23}  \\ && T_{23} \end{pmatrix}
        \begin{pmatrix} 1 \\ v_1 \\ v_2 \end{pmatrix} \\
        =
        T_{11} + T_{12} v_1 + T_{13} v_2 + T_{22} v_1^2 + T_{23} v_1 v_2 + T_{22} v_2^2
    \end{multline*}
    is a quadratic polynomial, with $\uuT$ being an upper triangular matrix.
\end{example}

\subsection*{Exact averaging of multilinear potentials for arbitrary body-orders}

In this work, we assume that the total energy is self-averaging in the sense of Lifshitz \cite{lifshitz_introduction_1988}, i.e., the energy, and derived properties depending on it, like elastic constants, cohesive energies, or defect energies, can, in principle, be averaged by considering a sufficiently large sample size.
For a given configuration, we are interested in computing the \emph{concentration average} $\Exp{\Pi}$ and, hence, treat the $s^i$'s as \emph{uncorrelated random variables} with probability $P(s^i = s^X) = c_X$, where $X$ is one of the alloy species, and $c_X$ is the concentration of $X$;
additionally, we have that $\sum_X c_X = 1$.
From the definition of the total energy, it follows that averaging $\Pi$ is equivalent to averaging $\clE$.
Finding an exact analytical expression of $\Exp{\clE}$ is the first main contribution of the present work.

Taking the average of \eqref{eq:multilinear_potential}, we obtain
\begin{equation}\label{eq:average_multi_pot}
    \Exp{\clE}
    =
    \Exp{
    T_{\< d \>}
    \times
    \left( \bigotimes_{k=1}^d \uv^k \right)
    }
    =
    T_{\< d \>}
    \times
    \Exp{
    \left( \bigotimes_{k=1}^d \uv^k \right)
    }
    .
\end{equation}
Averaging over all central atom species is easy since all site occupancies are uncorrelated.
The main difficulty is to average over the neighborhood atom species because naively averaging $\clE$ as is scales exponentially with the body-order.

In order to avoid introducing additional technicalities that are not essential, we will assume in the following that the feature vectors are always products of sums over per-neighborhood contributions $v^{k,j}$ so that
\[
    \uv^k = \sum_j \uv^{k,j}(\{ \ur^{ij}, s^i, s^j \})
    .
\]

\begin{remark}
    This requires re-defining, e.g., constant entries of $\uv^k$.
    For example, for feature vectors of type \eqref{eq:example_feature_vector}, we define 
    \[
        v_{(000)} = \sum_j \frac{1}{M},
    \]
    in the following, where $M$ is the number of atoms in the neighborhood of the $i$-th atom.
\end{remark}

With the definition of $\uv$ above, the problem of averaging of per-atom energies of multilinear potentials \emph{reduces to the problem of computing higher-order (statistical) moments} of $\uv^k$.
A way to obtain analytical expressions of higher-order moments of sums is to expand the average in terms of disjoint sums.
To elucidate this procedure, assume for the moment that $d = 2$ and expand the tensor product $\uv^1 \otimes \uv^2$ as follows
\begin{multline*}
    \uv^1 \otimes \uv^2
    =
    \left( \sum_{j_1} \uv^{1,j_1} \right) \otimes \left( \sum_{j_2} \uv^{2,j_2} \right) \\
    =
    \sum_j \uv^{1,j} \otimes \uv^{2,j}
    +
    \sum_{j_1 \neq j_2} \uv^{1,j_1} \otimes \uv^{2,j_2}
    .
\end{multline*}
Since the species vectors are uncorrelated, we can take the average of each feature vector with respect to the neighborhood species $Y$ individually for the second term such that
\begin{multline}\label{eq:expansion_d=2}
    \Exp{ \uv^1 \otimes \uv^2 }_Y
    =
    \sum_j \Exp{ \uv^{1,j} \otimes \uv^{2,j} }_Y \\
    +
    \sum_{j_1 \neq j_2} \Exp{ \uv^{1,j_1} }_Y \otimes \Exp{ \uv^{2,j_2} }_Y
    .
\end{multline}
Proceeding similarly for $d = 3$, a product of three sums can be expanded as
\begin{multline*}
    \uv^1 \otimes \uv^2 \otimes \uv^3
    =
    \left( \sum_{j_1} \uv^{1,j_1} \right) \otimes \left( \sum_{j_2} \uv^{2,j_2} \right) \otimes \left( \sum_{j_3} \uv^{3,j_3} \right)
    \\
    =
    \sum_j \uv^{1,j} \otimes \uv^{2,j} \otimes \uv^{3,j}
    +
    \sum_{j_1 \neq j_2} \uv^{1,j_1} \otimes \uv^{2,j_1} \otimes \uv^{3,j_2} \\
    +
    \sum_{j_1 \neq j_2} \uv^{1,j_1} \otimes \uv^{2,j_2} \otimes \uv^{3,j_1}
    +
    \sum_{j_1 \neq j_2} \uv^{1,j_1} \otimes \uv^{2,j_2} \otimes \uv^{3,j_2} \\
    +
    \sum_{j_1 \neq j_2 \neq j_3} \uv^{1,j_1} \otimes \uv^{2,j_2} \otimes \uv^{3,j_3}
    .
\end{multline*}
Again, since the species vectors are uncorrelated, we can write its average as
\begin{multline}\label{eq:expansion_d=3}
    \Exp{ \uv^1 \otimes \uv^2 \otimes \uv^3 }_Y
    =
    \sum_j \Exp{ \uv^{1,j} \otimes \uv^{2,j} \otimes \uv^{3,j} }_Y \\
    +
    \sum_{j_1 \neq j_2} \Exp{ \uv^{1,j_1} \otimes \uv^{2,j_1} }_Y \otimes \Exp{ \uv^{3,j_2} }_Y \\
    +
    \sum_{j_1 \neq j_2} \operatorname{\tau}_{(123)} \left( \Exp{ \uv^{1,j_1} \otimes \uv^{3,j_1} }_Y \otimes \Exp{ \uv^{2,j_2} }_Y \right) \\
    +
    \sum_{j_1 \neq j_2} \Exp{ \uv^{1,j_1} }_Y  \otimes\Exp{ \uv^{2,j_2} \otimes \uv^{3,j_2} }_Y \\
    +
    \sum_{j_1 \neq j_2 \neq j_3} \Exp{ \uv^{1,j_1} }_Y \otimes \Exp{ \uv^{2,j_2} }_Y \otimes \Exp{ \uv^{3,j_3} }_Y
    ,
\end{multline}
where $\operatorname{\tau}_{(123)}$ in the third term on the right hand side is the braiding map that aligns the indices of the tensor with the other terms.

\begin{figure*}[hbt]
    \centering
    \includegraphics[width=0.85\textwidth]{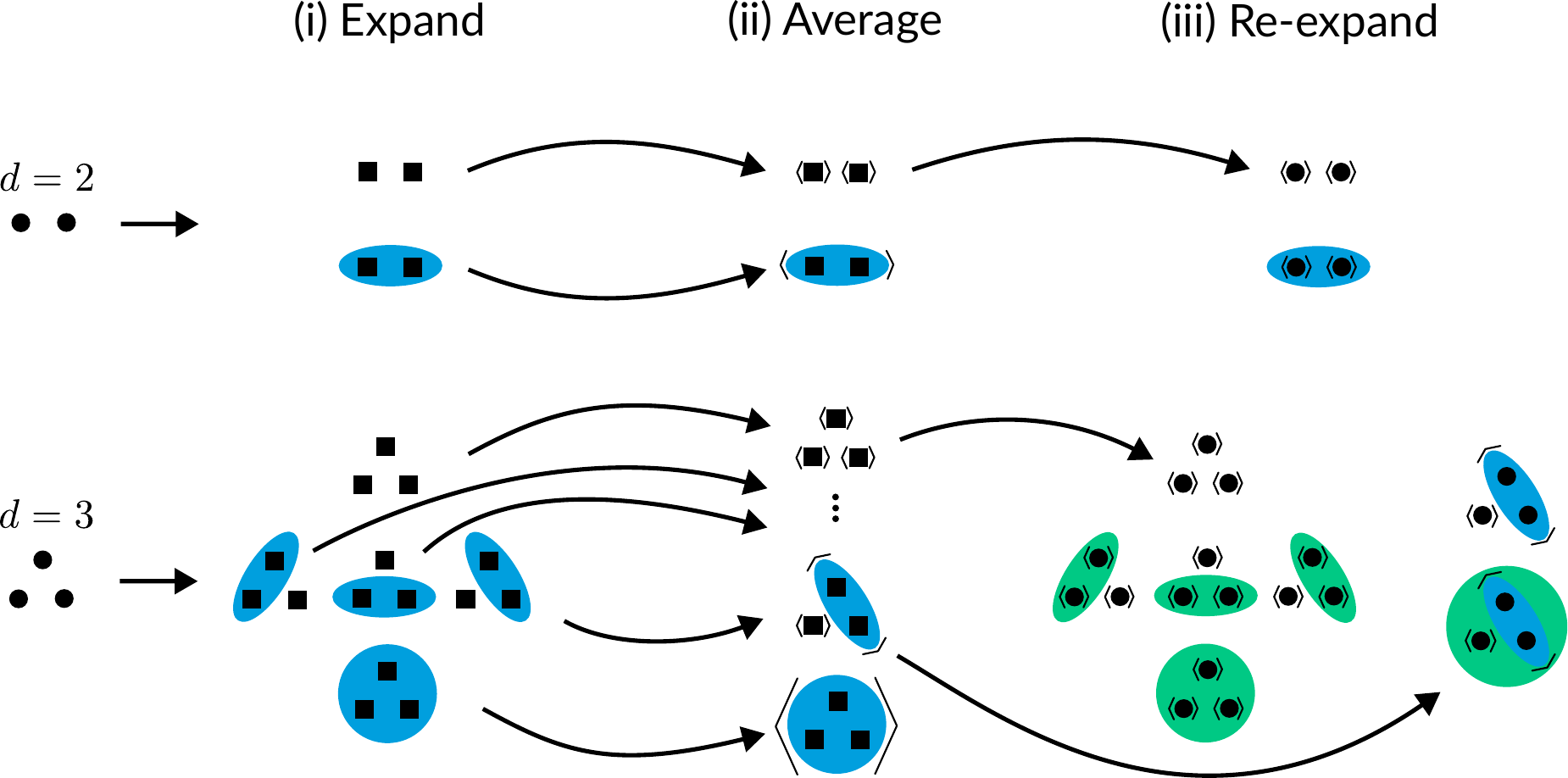}
    \caption{Schematic illustration of the procedure of set partitions for $d = 2$ and $d = 3$.
    (i) Expansion in terms of disjoint sums;
    sums with two or more elements are highlighted in blue and, for clarity, the elements of an expansion term are visualized with a square if the sums are disjoint, and a dot if not.
    (ii) Averaging of the disjoint sums.
    (iii) Re-expansion of all terms from (ii) in terms of products of sums;
    sums with two or more elements are highlighted in green}
    \label{fig:averaging_procedure}
\end{figure*}

\begin{example}[Average of the feature vector \eqref{eq:example_feature_vector}]
    The average of such a feature vector is given by
    \[
        \Exp{v_k}
        =
        \left\{
        \begin{aligned}
            1 & \quad k = 0, \\
            \sum_{X,Y} c_X c_Y \sum_j Q_\alpha(\vert \ur^{ij} \vert) z_\beta^X z_\gamma^Y Y_{\ell m}(\widehat\ur^{ij}) & \quad \text{else.}
        \end{aligned}
        \right.
        .
    \]
\end{example}

In principle, this procedure translates verbatim to higher body-orders, but, however, already becomes tedious from $d \ge 4$, especially in view of the need for re-expanding the disjoint sums in \eqref{eq:expansion_d=2} and \eqref{eq:expansion_d=3} in terms of products of sums in order to avoid the exponential scaling with the number of sums per term.
Therefore, it is desirable to use an automated procedure for arbitrary body-orders.

To that end, we use the procedure of set partitions as outlined, e.g., by \citet{mccullagh_tensor_2018} (Section 3.6), which is commonly used in statistics to derive analytical expressions of higher-order correlations.
This procedure consists of three steps, visualized in Figure \ref{fig:averaging_procedure}.
In step (i), we expand the tensor product $\uv^1 \otimes \ldots \otimes \uv^d$ into terms of disjoint sums, as done above.
The set of all expansion terms is called a \emph{partition}.
Each expansion term $p$ consists of blocks $b$, each representing a sum of outer products of $v$'s over all neighborhood atoms.
In step (ii), we take the average over all per-neighborhood-atom species for each of the (disjoint) sums---we can do this under the assumption made above that the site occupancies are uncorrelated.
In the final step, step (iii), we re-expand each of the expansion terms from step (ii) that consist of two or more disjoint sums into terms of products of sums.
An expansion of $p$ is again a partition with terms $\widetilde p$ consisting of blocks $\widetilde b$, each representing a sum of outer products of averages (of outer products of $v$'s) over all neighborhood atoms.
From Figure \ref{fig:averaging_procedure}, it can now be seen that the blocks $b$ are in fact subblocks of $\widetilde b$.
Finally, we also need to take the average over all per-central-atom species.
From the derivation above, a generic expression of the average potential energy can then be written as
\begin{equation}\label{eq:average_multilinear_pot}
\boxed{
    \Exp{\clE}
    =
    \sum_{p,\widetilde p}
    f_{p \widetilde p}
    \,
    \operatorname{\tau}_{p \widetilde p}(T_{\< d \>})
    \times
    \Exp{
    \bigotimes_{\widetilde b \in \widetilde p}
    \sum_j
    \bigotimes_{b \in \widetilde b}
    \Exp{\bigotimes_{l \in b} \uv^{l,j}}_Y
    }_X
    ,
}
\end{equation}
where $f_{p \widetilde p}$ is some integer prefactor that accounts for how often an expansion term is added or subtracted, and $\operatorname{\tau}_{p \widetilde p}$ braids the indices of $T_{\< d \>}$ to align them with the indices of the tensor products.

This procedure applies for any $d$ and thus provides a way to construct exact average potentials for arbitrary body-order in an automated fashion.
In the Supplementary Section \ref{sec:set_partitions}, we present a rigorous derivation of \eqref{eq:average_multilinear_pot} and an algorithm that automatically computes $p$, $\widetilde p$, and $f_{p \widetilde p}$.

At this point, we emphasize that the possibility of mixing the average model with the true random alloy species, which is relevant for predicting the standard deviations \eqref{eq:standard_deviations}, is included in \eqref{eq:average_multilinear_pot}.
It can simply be done by not averaging with respect to the central atom species $X$ or its neighboring atomic species $Y$.

\subsection*{Average two-body and three-body potentials}

For two-body potentials ($d = 1$), there is only one term, so, the average energy can directly be given
\[
    \Exp{\clE}
    =
    T_{\< d \>}
    \times
    \left(
    \sum_j \Exp{\uv^{j}}
    \right)
    .
\]

\begin{figure*}[hbt]
    \centering
    \includegraphics[width=\textwidth]{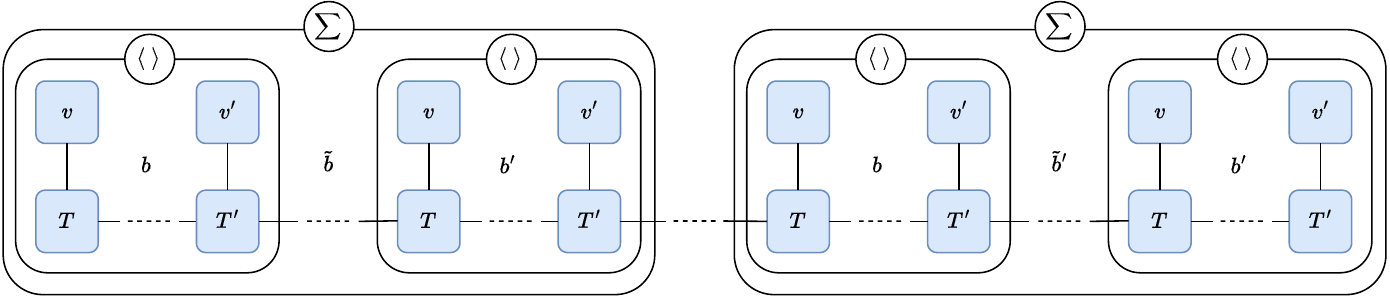}
    \caption{Tensor network diagram representation of some arbitrary term of the expansion \eqref{eq:average_multi_pot} using ETN potentials.
    In this diagram, a tensor is represented as a square and the tensor order is identified by the number of links that are attached to the square.
    Contractions over the tensors' dimensions can then be realized with connections between
    the squares.
    A more detailed description of this notation can be found in the Methods Section.
    At the first level, level 1, we contract all $T$'s with the $v$'s that correspond to a common block $b$ to order-one and order-two tensors and take the averages, visualized in the diagram with the $\< \, \>$ operator.
    At the second level, level 2, we contract those order-one and order-two tensors that correspond to a common block $\widetilde b$ to, again, order-one and order-two tensors;
    this is done for each neighborhood contribution, visualized with the $\sum$ operator.
    At the third level, level 3, we contract all order-one and order-two tensors computed at level 2 to a scalar.
    }
    \label{fig:average_etn_aligned}
\end{figure*}

For three-body potentials ($d = 2$), there is one term in \eqref{eq:expansion_d=2} consisting of two disjoint sums.
Taking the average and re-expanding it in terms of products of sums, we get
\begin{multline*}
    \sum_{j_1 \neq j_2} \Exp{ \uv^{1,j_1} }_Y \otimes \Exp{ \uv^{2,j_2} }_Y \\
    =
    \left( \sum_{j_1} \Exp{ \uv^{1,j_1} }_Y \right) \otimes \left( \sum_{j_2} \Exp{ \uv^{2,j_2} }_Y \right) \\
    -
    \sum_j \Exp{ \uv^{1,j} }_Y \otimes \Exp{ \uv^{2,j} }_Y
    .
\end{multline*}
The average energy of a three-body potential thus reads
\begin{multline*}
    \Exp{\clE}
    =
    T_{\< d \>}
    \times
    \Bigg\<
    \sum_j \Exp{ \uv^{1,j} \otimes \uv^{2,j} }_Y \\
    +
    \left( \sum_{j_1} \Exp{ \uv^{1,j_1} }_Y \right) \otimes \left( \sum_{j_2} \Exp{ \uv^{2,j_2} }_Y \right) \\
    -
    \sum_j \Exp{ \uv^{1,j} }_Y \otimes \Exp{ \uv^{2,j} }_Y
    \Bigg\>_X
    .
\end{multline*}

\subsection*{Implementation using equivariant tensor network potentials}

We now present an implementation of the average interatomic interaction model \eqref{eq:average_multilinear_pot} using equivariant tensor network (ETN) potentials.
The reasoning behind our choice is that we have to contract higher-order tensors like
\[
    S_{\< d \>}
    =
    \sum_j
    \Exp{
    \bigotimes_{k=1}^d \uv^{k,j}
    }_Y
\]
for the self-interaction terms (cf., the first terms in \eqref{eq:expansion_d=2} and \eqref{eq:expansion_d=3}).
However, the complexity of building such a $d$-dimensional tensor $S_{\< d \>}$ and contracting it with the coefficient tensor $T_{\< d \>}$, is proportional to $\bar{N}^d$, where $\bar{N}$ is the average size over all $v$'s.
With ETN potentials, we can avoid the exponential scaling with $d$ by factorizing the coefficient tensor $T_{\langle d \rangle}$ into order-two and order-three tensors $\uuT, \uuuT$, and contracting them with the $v$'s, thus avoiding building $S_{\< d \>}$ explicitly.

A possible decomposition of $T_{\< d \>}$ would be, e.g., the tensor train factorization \cite{oseledets_tensortrain_2011}%
\footnote{In quantum physics, this representation is known under the name matrix product state (e.g., \cite{perez-garcia_matrix_2007})}
\[
    T_{\< d \>} = \uuT^1 \times_2^1 \uuuT^2 \times_3^1 \ldots \times_3^1 \uuT^d,
\]
where $\times_2^1, \times_3^1$ now mean that we contract the last dimension of $\uuT^1, \ldots, \uuuT^{d-1}$ with the first dimension of $\uuuT^2, \ldots, \uuT^d$.
The order-two tensors $\uuT^1,\uuT^d$ are of size $N_1 \times r_1$ and $r_{d-1} \times N_d$, respectively, and the order-three tensors $\uuuT^k$ are of size $r_{k-1} \times N_k \times r_k$, where $r_1, \ldots, r_{d-1}$ are the ranks of the tensor network.
A multilinear form \eqref{eq:multilinear_potential} can then be realized by contractions with multiple order-two and order-three tensor as follows
\begin{multline*}
    \clE
    =
    \left( \uuT^1 \times_2^1 \left(\ldots \left( \uuuT^{d-1} \times_3^1 \left(\uuT^d \times_2^1 \uv^d\right) \right) \times_2^1 \uv^{d-1} \ldots\right) \right) \\ \times_1^1 \uv^1
    .
\end{multline*}
We refer to the Methods Section for a more detailed explanation of ETN potentials.

With such a representation of the per-atom energy, we are able to move the $T$'s inside the sums, and contract them with the feature vectors before taking the average
\begin{multline*}
    T_{\langle d \rangle}
    \times
    \left(
    \sum_j
    \Exp{
    \bigotimes_{k=1}^d \uv^{k,j}
    }_Y
    \right)
    =
    \sum_j
    \bigg\< \\
    \left( \uuT^1 \times_2^1 \left(\ldots \left( \uuuT^{d-1} \times_3^1 \left(\uuT^d \times_2^1 \uv^{d,j}\right) \right) \times_2^1\uv^{d-1,j} \ldots\right) \right) \\
    \times_1^1 \uv^{1,j}
    \bigg\>_Y
    .
\end{multline*}
The complexity of this operation is $d \bar{r}^2 \bar{N}$, where $\bar{r}$ is the average rank of the tensor network, which is, obviously, much more efficient than contracting $T_{\langle d \rangle}$ with $S_{\langle d \rangle}$---provided that the ranks can be kept small.

Figure \ref{fig:average_etn_aligned} visualizes the implementation of \eqref{eq:average_multilinear_pot} using ETN potentials in terms of tensor network diagrams;
for a detailed derivation of this representation, the reader is referred to the Supplementary Section \ref{sec:algo}.
From this tensor network diagram, it can be immediately deduced that the only operations that are required to compute $\Exp{\clE}$ are contractions of order-three tensors with one or two vectors.
This simplifies the automation of computing average energies using potentials with higher body-orders, as well as the automation of differentiating $\Exp{\clE}$ in order to compute average forces, stresses, etc.
This is the second main contribution of the present work.

\subsection*{Comparison with Monte Carlo sampling}

We now validate the averaging formalism by demonstrating the convergence of Monte Carlo sampling using random supercells with a growing number of atoms to the average alloy \eqref{eq:average_multilinear_pot}.
We benchmark the convergence on various bulk and defect energies.
Details on how we compute those properties can be found in the Methods Section under "Simulation details".

To that end, we first construct a four-body ETN potential by training it on the training set of \citet{li_complex_2020} for the NbMoTaW medium-entropy alloy.
The four-body SNAPs \cite{li_complex_2020} and MTPs \cite{yin_atomistic_2021} trained on this training set have been shown to be able to capture properties relevant for predicting mechanical properties, e.g., lattice constants, elastic constants, etc., as well as crucial defect properties, such as stacking fault energies, and dislocation core structures, and is thus ideally suited for validation purposes.
The corresponding ETN training errors, shown in Table \ref{tab:training_errors}, are lower than those reported in \cite{li_complex_2020} for the SNAP and, therefore, our ETN potential should be at least as accurate as the SNAP in capturing bulk and defect properties.
Details on the training can be found in the Methods Section under "ETN training".
The functional form of an average four-body ETN potential is given in the Supplementary Section \ref{sec:four_body}.

\begin{table}[hbt]
    \centering
    \begin{tabular}{|c|c|c|}
        \hline
        \backslashbox{Quantity}{Potential} & \multicolumn{1}{|c|}{SNAP} & \multicolumn{1}{|c|}{ETN} \\ \hline\hline
        Energies [meV/atom] & 4.30 & 3.58 \\ \hline\hline
        Forces [meV/\AA] & 0.130 & 0.069 \\ \hline\hline
        Stresses [eV] & N/A & 0.48 \\ \hline
    \end{tabular}
    \caption{Mean absolute training errors for the SNAP from \cite{li_complex_2020} and the ETN potential;
    stress errors for the SNAP were not given in \cite{li_complex_2020}.
    The errors for the ETN potential are smaller, indicating that it should be at least as good as the SNAP for predicting bulk and defect properties}
    \label{tab:training_errors}
\end{table}

\begin{figure}[hbt]
    \centering
    \includegraphics[width=0.97\linewidth]{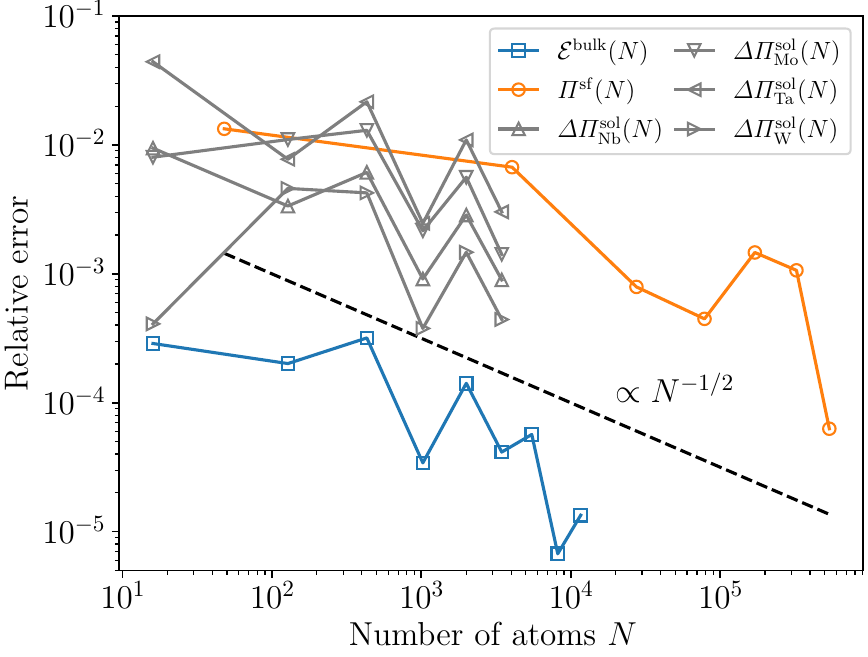}
    \caption{Relative errors of the bulk cohesive energy, the stacking fault energy, and the solute energies, computed with random configurations as a function of the number of atoms per supercell with the respect to the average ETN potential}
    \label{fig:mc_error}
\end{figure}

We first test the convergence of the bulk cohesive energy.
As expected, the average cohesive energy in the random alloy converges to the average ETN value with the usual Monte Carlo convergence rate of $N^{-1/2}$, as shown in Figure \ref{fig:mc_error}, providing evidence for the correctness of our implementation of the average model.

Next, we test the convergence of the $1/4[111]$ stacking fault energy.
Compared to the cohesive energies, the convergence rate is the same, as expected, providing further validation of our average model (cf., Figure \ref{fig:mc_error}).
Moreover, we point out that the relative error is about 1.5--2 times higher for the stacking fault energies---this highlights the greater complexity of averaging defect energies.

We further test our average model for the interaction energies of the average alloy with the (true) alloy species.
This can be implemented within our average model by not averaging over $X$, or $Y$, depending whether the solute sits at the central or a neighboring site.
The alloy species are effectively solutes in the average matrix, so we call those interaction energies "solute energies" in the following.
Solute energies play an important role in computing deviations from average random alloy properties (cf., \cite{nohring_design_2020}).
As shown in Figure \ref{fig:mc_error}, the solute energies computed in the random alloy converge to the solute energies computed using the average model.
Note that the prefactor is more than an order of magnitude higher than for the cohesive energies, comparable to the stacking fault energies, showing that solute energies are more difficult to average via random sampling.
This demonstrates that average ETN potentials \emph{\textbf{exactly} capture deviations of random alloy properties from the average}.

\subsection*{Comparison with average EAM potentials}

\begin{table*}[hbt]
    \centering
    \begin{tabular}{|c|l|l|l|l|}
        \hline
        \backslashbox{Method}{Solute} & \multicolumn{1}{|c|}{Nb} & \multicolumn{1}{|c|}{Mo} & \multicolumn{1}{|c|}{Ta} & \multicolumn{1}{|c|}{W} \\ \hline\hline
        DFT & 1.153 & -1.251 & 1.132 & -1.034 \\ \hline\hline
        EAM & 1.019 (12\,\%) & -1.218 (3\,\%) & 1.181 (4\,\%) & -0.845 (18\,\%) \\ \hline\hline
        ETN & 1.105 (4\,\%) & -1.237 (1\,\%) & 1.105 (2\,\%) & -0.979 (5\,\%) \\ \hline
    \end{tabular}
    \caption{Misfit volumes computed with DFT, average EAM potentials, and average ETN potentials.
    The agreement between DFT and the ETN is very good, while the EAM potential shows consistently larger errors than the ETN.
    The DFT and EAM values are taken from \cite{maresca_mechanistic_2020}}
    \label{tab:misfit_volumes}
\end{table*}

\begin{figure*}[hbt!]
    \centering
    \includegraphics[width=0.7\textwidth]{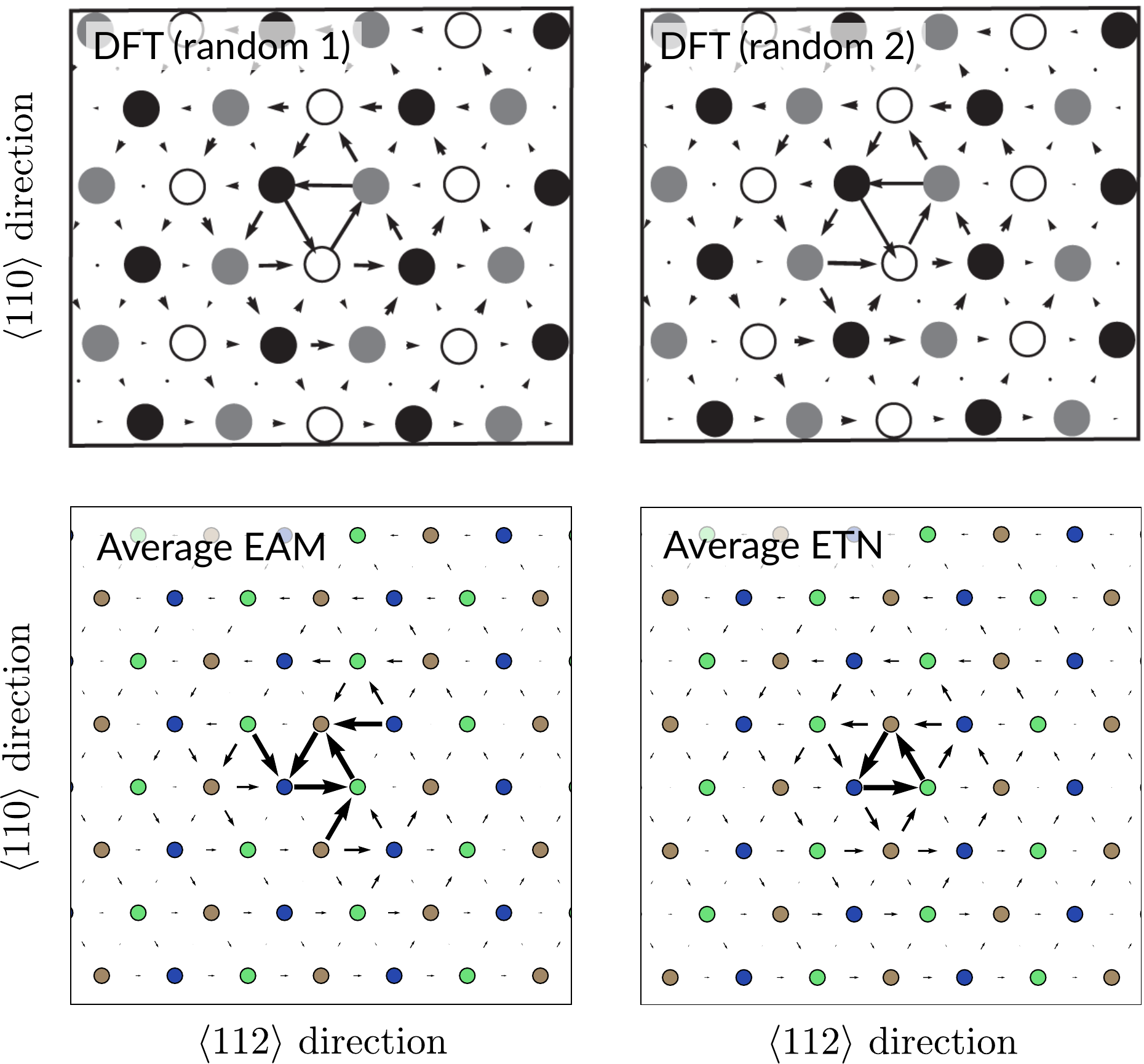}
    \caption{Differential displacement map \cite{vitek_theory_1974} of the relaxed $1/2 \left\langle 111 \right\rangle$ screw dislocation core in NbMoTaW using DFT for two different random configurations, and using the average EAM and ETN potentials;
    The DFT cores are reversed because they have been extracted from a dipole configuration, but physically all cores are equivalent.
    The EAM potential converges to an artificial polarized core configuration, while the ETN core is perfectly symmetric as expected from the random DFT configurations.
    The DFT results are reproduced from \cite{yin_initio_2020} (Figure 1) with permission}
    \label{fig:MoNbTaW_diff_displ}
\end{figure*}

Given the successful validation of the average model from the previous section, we now turn to properties that are relevant for calibrating strengthening models for solid solutions, namely misfit volumes and dislocation core structures \cite{varvenne_theory_2016,maresca_theory_2020,maresca_mechanistic_2020}.
Previously, those properties have been computed with averaged embedded atom method (EAM) potentials of \citet{zhou_misfitenergyincreasing_2004} (cf., Methods Section, "Average EAM potential"), but there are many compositions of the NbMoTaW alloy, in particular those having a Mo concentration of more than 10\,\%, for which the EAM potential is neither qualitatively---in the case of misfit volumes---nor qualitatively---the EAM potential predicts artificial polarized core structures---accurate.
We show in the following that average \emph{ETN potentials are able to capture those properties}.

In Table \ref{tab:misfit_volumes}, we show the misfit volumes computed with the average ETN potential, as well as those computed with the average EAM, and DFT.
While the EAM potential is accurate for Mo and Ta, the errors for Nb and W are much higher, between 10 and 20\,\%; errors of this size have been considered to be too large for making predictions on real alloys \cite{maresca_mechanistic_2020}.
On the other hand, for the average ETN potential, the deviation from the DFT results is at most 5\,\%.
This demonstrates that average ETN potentials are capable of reliably predicting misfit volumes and can therefore be considered as a more efficient alternative to DFT for calibrating models of solid-solution strengthening.

\begin{figure*}[hbt]
    \centering
    \includegraphics[width=0.7\textwidth]{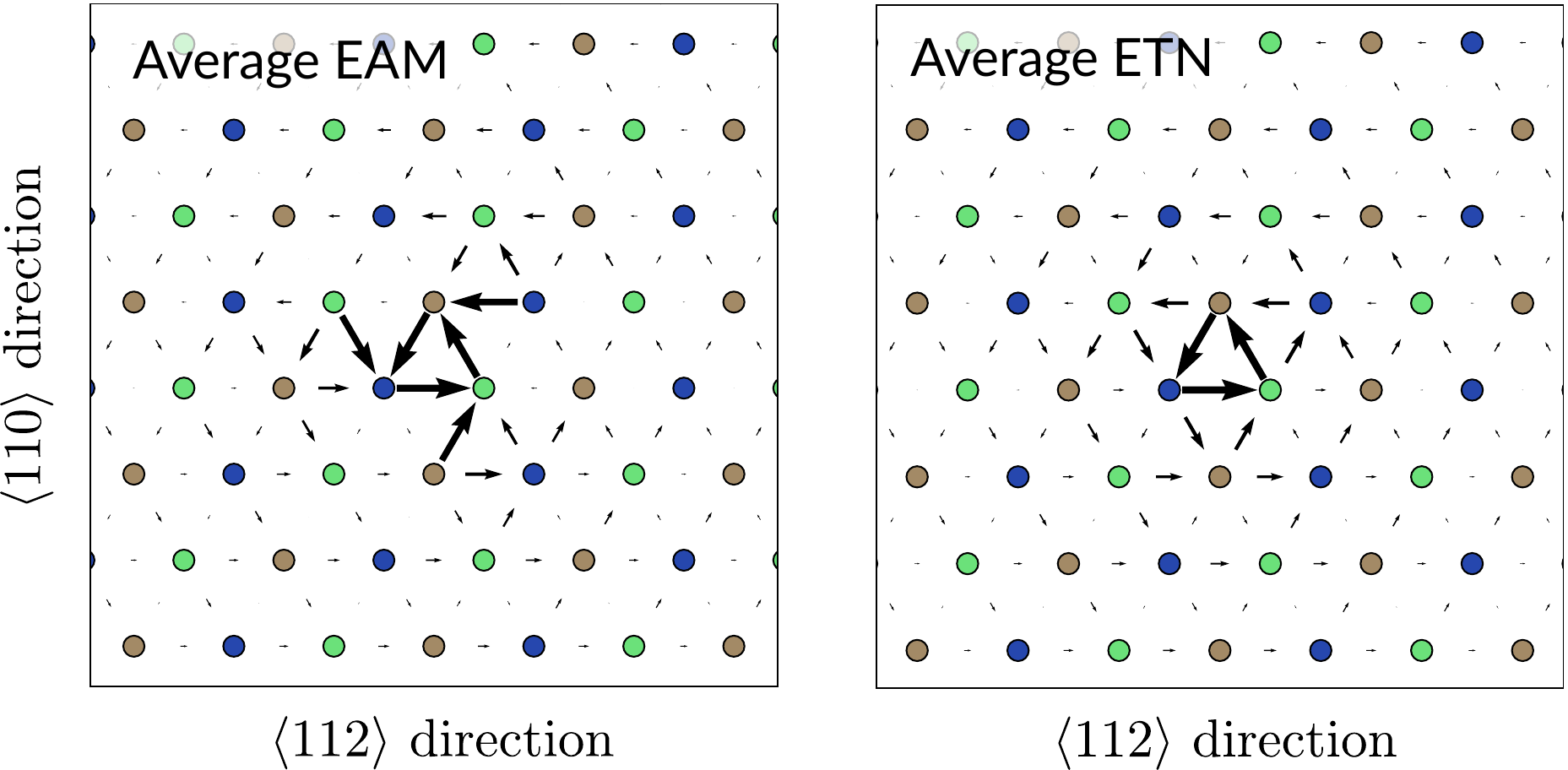}
    \caption{Differential displacement map \cite{vitek_theory_1974} of the relaxed $1/2 \left\langle 111 \right\rangle$ screw dislocation core in NbMo using DFT for two different random configurations, and using the average EAM and ETN potentials.
    The EAM potential converges to an artificial polarized core configuration, while the ETN core is perfectly symmetric as expected from the random DFT configurations (cf., Figure \ref{fig:MoNbTaW_diff_displ})}
    \label{fig:MoNb_diff_displ}
\end{figure*}

We now validate the average ETN potential by computing the core structure for the $1/2 \left\langle 111 \right\rangle$ screw dislocations.
More precisely, we compute the structure of the so-called easy-cores because of the deficiency of EAM potentials predicting polarized cores in refractory alloys from the NbMoTaWV family for certain compositions, in particular those having higher Mo content of $>$\,10\,\% \cite{maresca_theory_2020}.
Polarized cores have not been observed in random realizations of such alloys. 
Existing DFT results \cite{yin_initio_2020} show an essentially compact six-fold symmetric core, with some small asymmetries that can be attributed to the local randomness around the core, as shown in Figure \ref{fig:MoNbTaW_diff_displ};
but these asymmetries should cancel out when taking the average over a sufficient large number of samples.
Other authors \cite{li_complex_2020} have computed the core structure with the DFT-accurate SNAP for different random configurations and obtained very similar results.

Using this simulation setup, we obtain the polarized core structure for the average EAM potential in Figure \ref{fig:MoNbTaW_diff_displ}, which possesses a three-fold symmetry.
Using the average ETN potential, such a polarized configuration does not occur during relaxation, so, we obtain a perfectly symmetric non-polarized core structure.
In addition, we have computed the dislocation core structure in the equiatomic NbMo binary alloy that also suffers from polarized cores, as reported in \cite{maresca_theory_2020}.
Again, relaxing the configuration using the average ETN potential gives a perfectly symmetric core structure (cf., Figure \ref{fig:MoNb_diff_displ}).
Hence, the examples presented in this section show that the average ETN potential is able to predict a core structure consistent with the core structure that one expects from a random sampling over several core structures computed with DFT and other MLIPs.

\section*{Discussion}

In summary, we have developed a framework for averaging linear MLIPs that allows to make \textbf{DFT-accurate predictions of average random alloy properties and deviations from them} within the scope of the effective medium theory.
Our framework is based on the observation that linear MLIPs can be represented as multilinear forms, with the arguments being feature vectors that are sums over pair-wise interactions of the central atom with its neighboring atoms.
Then, using established tools from higher-order statistics and graph theory, we are able to fully automatically expand the average per-atom energy in terms of disjoint sums, take the analytical average, and and re-expand them in terms of products of sums so that the evaluation of the average per-atom energy scales linearly with the size of an atomic neighborhood.
In such an expansion, there appear higher-order tensors due to the averaging of self-interactions between feature vectors.
To avoid forming those higher-order tensors, we have developed an implementation using equivariant tensor network (ETN) potentials in which the self-interacting feature vectors are contracted before taking the average.
We have benchmarked our average model on the problem of predicting crucial properties of the NbMoTaW medium-entropy alloy and shown that the average ETN potential is able to overcome well-known deficiencies of state-of-the-art EAM potentials.
In particular, we have shown that the average ETN potential predicts a compact $1/2 \left\langle 111 \right\rangle$ screw dislocation core structure, consistent with DFT, whereas the EAM potential predicts an artificial polarized core structure.

An important feature of our averaging scheme is the possibility of combining the average potential with the original alloy species so that "real" atoms can be inserted into defects created in the "average" alloy.
Hence, we anticipate that our average potential can be applied to predict interaction energies of a dislocation in the average alloy with the real atoms, which enter into models of screw dislocation strengthening and cannot be computed with the average EAM potentials due to the above limitations (cf., \cite{baruffi_screw_2022}).
Another potential application could be segregation of real atoms to grain boundaries in the average alloy \cite{scheiber_initio_2015}.

More broadly, the methodology outlined here could potentially also be applied to other materials that form solid solutions, such as perovskites, or for averaging other quantities like spin orientations.

A shortcoming of the present implementation is the reduced efficiency compared to the ETN potential for the true random alloy since the computational cost scales quadratically with the number of species and is, moreover, proportional to the number of expansion terms.
In the following, we outline some potential improvements of the efficiency of average ETN potentials.

The quadratic scaling with respect to the number of species can, in principle, be overcome by contracting the tensors $T^1, \ldots, T^d$ with the species-related coefficients in a preprocessing step.
To reduce the number of expansion terms, one possibility is to expand the per-atom energy around the average feature vector $\bar\uv$, that is,
\begin{multline*}
    \clE
    =
    \clE(\uv)
    =
    \clE(\bar\uv)
    +
    \grad{\rm v}{\sT}{\clE(\bar\uv)}
    \cdot (\uv - \bar\uv) \\
    +
    \grad{\rm v}{2}{\clE(\bar\uv)}
    \times \Big( (\uv - \bar\uv) \otimes (\uv - \bar\uv) \Big)
    +
    \ldots
    .
\end{multline*}
Taking averages yields
\begin{multline*}
    \Exp{\clE(\uv)}
    =
    \clE(\bar\uv)
    +
    \grad{\rm v}{\sT}{\clE(\bar\uv)}
    \cdot \Exp{(\uv - \bar\uv)^\sT} \\
    +
    \grad{\rm v}{2}{\clE(\bar\uv)}
    \times \Exp{\Big( (\uv - \bar\uv) \otimes (\uv - \bar\uv) \Big)}
    +
    \ldots
    .
\end{multline*}
Since the absolute value of each expansion term can be bounded from above by $N_{\rm spec}^{-n}$, where $N_{\rm spec}$ is the number of species and $n$ is the order of the expansion term, it appears that a good approximation to $\Exp{\clE}$ could already be realized by considering only a few lower-order terms.
In fact, we have already implicitly confirmed this in a previous publication \cite{novikov_aiaccelerated_2022}, where we have constructed an average MTP by replacing the moment tensor descriptors with their averages, which corresponds to an expansion of $\Exp{\clE}$ to zeroth order.
Moreover, there are many repetitive operations occuring in the expansion \eqref{eq:average_multilinear_pot} (like taking the average of a single feature vector) whose number grows likewise with the number of expansion terms, which can also be exploited.
Another interesting direction could be replacing the pair-wise features with triplet features that would allow for a lower body-order (cf., \cite{nigam_completeness_2024}).

In order to construct an average ETN potential that is as efficient as the original (non-average) multi-component ETN potential, one could consider a transfer learning approach.
Conceptually, this can be achieved by training a \emph{new ETN potential} on energies, forces, and stresses, computed using the \emph{average ETN potential}, in the same way as MLIPs are typically trained on quantum-mechanical models using standard techniques, such as active learning (e.g., \cite{podryabinkin_active_2017,hodapp_operando_2020}).
Moreover, since the average ETN potential is a local model, it is easy to use arbitrary clusters of atoms as training configurations.
This heavily simplifies the training procedure since we are not limited to training on periodic configurations, as it is the case for plane-wave DFT.

Finally, we would like to remark that the present average model disregards short-range ordering that can have an influence on deformation mechanisms (cf., \cite{rasooli_deformation_2024}).
Short-range ordering adds another level of complexity to the problem of taking averages because one needs to take correlations between atomic species into account.
This would require a new set partitioning method that separates the clusters of atoms in which species are correlated, but perhaps remains tractable if the correlations can be approximated with polynomials of pair-wise correlation functions.

\section*{Methods}

\subsection*{Equivariant tensor network potentials}

\paragraph*{Symbolic representation}

Tensor networks are a way of representing high-dimensional tensors in a low-rank format by factorizing the full tensor into smaller tensors, up to the order of three.
There are different formats, with tensor trains, hierarchical Tucker, or PEPS, arguably being among the most popular ones (cf., \cite{cichocki_tensor_2017,orus_tensor_2019}), but their contractions with vectors $\uv^1$, $\uv^2$, etc., can all be realized by sequences of contractions of up-to-order-three tensors
\begin{align*}
    \uu^1 &= \uuT^1 \times_2^1 \uv^1, \\
    \uu^2 &= (\uuuT^2 \times_3^1 \uv^2) \times_2^1 \uu^1, \\
    \uu^3 &= (\uuuT^3 \times_3^1 \uv^3) \times_2^1 \uu^2, \\
          &\ldots
          ,
\end{align*}
where $\times_2^1,\times_3^1$ imply that we contract the second, or third dimension, respectively, of the $T$'s with the first dimension of the operand.
The main difference of ETNs compared to conventional tensor networks is the implementation of symmetry constraints that render the ETN invariant under actions of the corresponding symmetry group.
In our case of interatomic potentials, we require that the per-atom energy $\clE$ stays invariant under actions of the group of rotations SO(3).%
\footnote{To encode the full O(3) invariance (rotations and reflections) into ETNs, it suffices that $\clE$ is a real quantity (cf., \cite{hodapp_equivariant_2023}, section 3.4.2)}
To that end, we require the feature vectors $\uv$ to be (SO(3))-covariant vectors, that rotate correspondingly with a basis change under SO(3).
In the following we consider a decomposition of $\uv$ into an irreducible covariant representation of SO(3) using spherical harmonics.
Thus, we consider $\uv$ as a multi-index vector $v_{(\ell mn)}$, with $\ell = 0, \ldots, L$ being  the index of the subspace of the irreducible representation, $m \in \{ -\ell, -\ell + 1, \ldots, \ell \}$ is the dimension of the subspace, and $n = 1, \ldots, N(\ell)$ is the number of radial channels corresponding to each $\ell$.
We point out that we intentionally deviate from the ordering $n\ell m$, commonly used in quantum physics, that puts the index $n$ in front of $\ell$ and $m$.
We have chosen this notation because each multi-index in our tensor network may depend on a different $\ell$, so, since $m$ and $n$ are always assumed to depend on $\ell$, the ordering $\ell m n$ appears to be more comprehensible in our context.

With this definition of the feature vectors, a sufficient condition for $\clE$ being invariant under actions of SO(3) is that the tensors $\uuuT$ are equivariant maps of those covariant vectors.
According to the Wigner-Eckhart Theorem, any $\uuuT$ with three multi-indices $\{ (\ell_i, m_i, n_i) \}_{i=1,\ldots,3}$, can be factorized to
\begin{multline*}
    T_{(\ell_1 m_1 n_1)(\ell_2 m_2 n_2)(\ell_3 m_2 n_3)} \\
    =
    \mtheta_{(\ell_1 n_1)(\ell_2 n_2)(\ell_3 n_3)}
    C_{(\ell_1 m_1)(\ell_2 m_2)(\ell_3 m_3)}
    ,
\end{multline*}
where $\mtheta_{(\ell_1 n_1)(\ell_2 n_2)(\ell_3 n_3)}$ is the tensor of model coefficients, and $C_{(\ell_1 m_1)(\ell_2 m_2)(\ell_3 m_3)}$ is the Clebsch-Gordan coefficient that defines the symmetry group.
As a tensor network, we use an (equivariant) tensor train representation \cite{oseledets_tensortrain_2011} of atomic energies with equal feature vectors $\uv^1, \ldots, \uv^d = \uv$.
An ETN potential in the tensor train format can then be written as follows
\begin{multline}\label{eq:etn_pot}
    \clE
    =
    \left(T^1_{(\ell_1' m_1' n_1') (\ell_1 m_1 n_1)} v_{(\ell_1' m_1' n_1')}\right) \\
    \left(T^2_{(\ell_1 m_1 n_1) (\ell_2' m_2' n_2') (\ell_2 m_2 n_2)} v_{(\ell_2' m_2' n_2')}\right) \\
    \ldots
    \left(T^d_{(\ell_{d-1} m_{d-1} n_{d-1}) (\ell_d' m_d' n_d')} v_{(\ell_d' m_d' n_d')}\right)
    .
\end{multline}
in which the \emph{channels} $n_1,n_2,\ldots,n_d$ \emph{naturally emerge as ranks of the tensor network}.
As feature vectors, we use the ones given in \eqref{eq:example_feature_vector}, with a further contraction of the radial features before entering the tensor train as follows
\[
    v_{(\ell m n)}
    =
    \sum_j \Big( B_{\ell n\alpha\lambda} Q_\alpha(\vert \ur^{ij} \vert) \big( A_{\ell\lambda\beta\gamma} z_\beta^i z_\gamma^j \big) \Big) Y_{\ell m}(\widehat\ur^{ij})
\]
to avoid the problem of an exponentially growing size of the feature vectors (cf., \cite{hodapp_equivariant_2023}).
This enables learning similarities between \emph{all} radial features from the data through the parameter tensors $A$ and $B$.
The number of coefficients of ETN potentials is proportional to $d \bar{r}^2 \bar{n}$, where $\bar{r}$ is average rank of the tensor network, and $\bar{n}$ is the average size over all dimensions of the feature vector.
This makes them obviously much more efficient than the raw polynomial representation \eqref{eq:multilinear_potential} if $\bar{r}$ is small;
in practice they require, e.g., two to three times fewer coefficients than the multi-component MTPs of \citet{gubaev_accelerating_2019} that use some semi-empirical feature compression.

\paragraph*{Tensor network diagram representation}

It is difficult to write ETN potentials using mathematical formulas, and in doing so is even more difficult for the expansion terms \eqref{eq:average_multilinear_pot}.
To that end, we have introduced a graphical representation inspired by tensor diagrams from quantum physics \cite{bridgeman_handwaving_2017} in which tensors are represented as square blocks, e.g., for a vector $\uv$, a matrix $\uuA$, and an order-three tensor $\uuuT$, we may equivalently write
\[
    \begin{aligned}
        \includegraphics[scale=0.8]{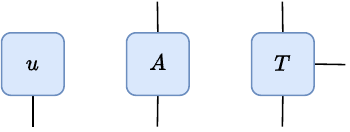}
    \end{aligned}
    ,
\]
with the tensor order being identified by the number of links that are attached to it.
Contractions over the tensors' dimensions can then be realized with connections between the blocks as follows
\[
    \begin{aligned}
        \includegraphics[scale=0.8]{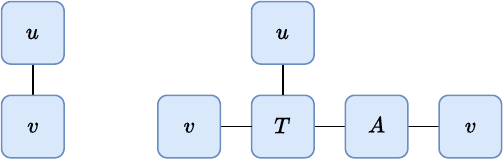}
    \end{aligned}
    .
\]
With such a diagrammatic notation, the ETN potential \eqref{eq:etn_pot} can be written as follows
\[
    \clE =
    \begin{aligned}
        \includegraphics[scale=0.8]{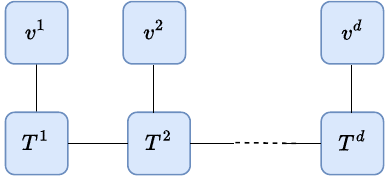}
    \end{aligned}
    .
\]
It is easy to appreciate the benefits of this notation because one can directly deduce the structure of the potential and the order of the involved tensors.
It is then straightforward to visualize the operations involved in computing an expansion term of \eqref{eq:average_multi_pot} using ETN potentials.

\subsection*{ETN training}

For our four-body ETN potential, we use the best set of hyperparameters from \cite{hodapp_equivariant_2023} that were found using a greedy structure search (cf., Figure 10 in \cite{hodapp_equivariant_2023}, last iteration).
Further, we choose a cut-off radius of 5\,\AA.
The training set contains 5\,529 configurations in total that have been calculated with density functional theory (DFT), including various types of configurations, e.g., configurations containing free surfaces, snapshots from finite temperature ab initio molecular dynamics simulations, etc.
We fit to total energies $\Pi$, forces $\uf$, and stresses $\uusigma$, by minimizing the loss function
\begin{multline*}
    \scL =
    \sum_i \Bigg(
	w_{\rm e} \Big( \Pi^{\rm etn}_i - \Pi^{\rm dft}_i \Big)^2
	+
	w_{\rm f} \left( \sum_j \| \uf^{\rm etn}_{i,j} - \uf^{\rm dft}_{i,j} \|^2 \right) \\
    +
    w_{\rm s} \| \uusigma^{\rm etn}_i - \uusigma^{\rm dft}_i \|^2
    \Bigg)
    ,
\end{multline*}
with the weights set to
\begin{align*}
    w_{\rm e} = 1 \, {\rm eV^2},
    &&
    w_{\rm f} = 0.01 \, {\rm (eV/\text{\AA})^2},
    &&
    w_{\rm s} = 0.001 \, {\rm eV^2}
    .
\end{align*}
To minimize the loss function, we have used \texttt{SciPy}'s BFGS solver.
We have terminated the minimization after 3\,000 iterations.

\subsection*{Simulation details}

\paragraph*{Ground state}

To compute the material's ground state, we first minimize the total energy of the primitive cell with respect to the lattice constant using the Nelder-Mead simplex algorithm as implemented in \texttt{SciPy}.
We consider a configuration as converged when the difference between the solutions for two subsequent iterations is less than $10^{-10}$\,\AA.
We have compared the results of the average alloy with a random supercell of 2\,000 atoms and the results for the lattice constant are identical, 3.24\,\AA, which matches the DFT result reported in \cite{maresca_mechanistic_2020}.

\paragraph*{Cohesive energy}

In the average alloy this is simply the per-atom energy $\clE^{\rm bulk}$ of the ideal bcc lattice.
To compute the average cohesive energy in the random alloy, we use supercells with increasing size, compute the total energy, and divide it by the number of atoms in the cell.

\paragraph*{\texorpdfstring{$1/4[111]$}{1/4[111]} stacking fault energy}

We create a rectangular prismatic supercell configuration with an orientation of the axes given by $[11\overline{2}]$, $[\overline{1}10]$, and $[111]$, respectively.
We denote this configuration by $\{ \ur^i \}_{\rm bulk}$.
We then translate half of the crystal by one Burgers vector in the $[111]$ direction to create the configuration with the stacking fault that we denote in the following by $\{ \ur^i \}_{\rm sf}$.
We further apply a shear displacement of half a Burgers vector to the cell vectors so that there will be only one stacking fault per supercell, and every atom at the periodic boundary sees a perfect crystalline environment.
The stacking fault energy is then the difference between both configurations divided by the area of the slip plane $A$
\[
    \Pi^{\rm sf} = \frac{\Pi^{\rm sf}(\{ \ur^i \}_{\rm sf}) - \Pi^{\rm sf}(\{ \ur^i \}_{\rm bulk})}{A}
    .
\]
Again, To compute the average stacking fault energy in the random alloy, we use supercells with increasing size.

\paragraph*{Solute energies}

In the average alloy, a solute energy is computed by taking the energy difference between two configurations, one with a solute $X$, and one without the solute, i.e.,
\[
    \Delta\Pi^{\rm sol}_X = \Pi^{\rm sol}_X - \Pi^{\rm bulk}
    .
\]
For computing solute energies using random configurations, we loop over all atoms in the configuration, replace each atomic species with the solute, compute the energy difference with respect to the bulk, and divide by the number of atoms as follows
\[
    \Delta\Pi^{\rm sol}_X(N) = \frac{\sum_{i=1}^N \Pi^{{\rm sol},i}_X(N) - \Pi^{\rm bulk}(N)}{N}
    ,
\]
where $\Pi^{{\rm sol},i}_X$ is the energy of a random configuration with the solute $X$ sitting at the $i$-th lattice site.

\paragraph*{Misfit volumes}

For computing the misfit volumes in the average alloy, we proceed in the same way as for the solute energies.
For the average alloy, we first compute the volume of the bulk configuration, and then replace any of the average species with a solute of type $X$.
We then relax the supercell with respect to the cell volume using the Nelder-Mead algorithm until the differences in the length of the cell between two subsequent iterations became less than $10^{-10}$\,\AA.
The misfit volume for a type $X$ solute is then given by
\[
    \Delta V^{\rm sol}_X = V^{\rm sol}_X - V^{\rm bulk}
    .
\]

\paragraph*{Dislocation core relaxation}

For computing the dislocations, we use a cylindrical configuration with a radius 35 times the magnitude of the Burgers vector containing $\sim$\,6\,000 atoms.
We fix the outermost atoms atoms up to a radial distance of two times the cut-off radius, so effectively there are $\sim$\,5\,000 free atoms in the simulation region.
Outside of the simulation region we fix the displacement of the atoms to the linear elastic solution of a screw dislocation;
the elastic constants for equiatomic MoNbTaW and MoNb that are necessary for setting up the boundary conditions agree reasonably well with DFT whereas the average EAM potential deviates from the DFT shear modulus by about 50\,\% (cf., Table \ref{tab:elastic_constants}).
The relaxation is then performed using the Fast Inertial Relaxation Engine \cite{bitzek_structural_2006}, as implemented in the Atomic Simulation Environment (ASE) \cite{hjorthlarsen_atomic_2017}.
A simulation is considered converged when the maximum force on an atom is less than $10^{-6}$\,eV/\AA;
this is the same tolerance as used by \citet{maresca_theory_2020}.

\begin{table}[hbt]
    \centering
    \begin{tabular}{|c|l|l|l|}
        \hline
        \backslashbox{Potential}{$C_{ij}$} & \multicolumn{1}{|c|}{$C_{11}$} & \multicolumn{1}{|c|}{$C_{12}$} & \multicolumn{1}{|c|}{$C_{44}$} \\ \hline\hline
        MoNbTaW (ETN) & 386 & 150 & 49 \\ \hline\hline
        MoNbTaW (EAM) & 348 & 173 & 96 \\ \hline\hline
        MoNbTaW (DFT) & 374 & 163 & 64 \\ \hline\hline
        MoNb (ETN)    & 363 & 141 & 45 \\ \hline
        MoNb (EAM)    & 342 & 160 & 75 \\ \hline
    \end{tabular}
    \caption{ETN and DFT elastic constants in GPa; the DFT values are taken from \cite{maresca_mechanistic_2020}}
    \label{tab:elastic_constants}
\end{table}

\subsection*{Average EAM potentials}

The general form of an EAM potential is given by (cf., e.g., \cite{daw_embeddedatom_1993})
\[
    \clE
    =
    \sum_j \mphi(\ur^{ij},s^i,s^j) + \sum_{j} F(\mrho(\ur^{ij},s^i),s^i),
\]
where $\mphi$ is a pair potential, and $F$ is the embedding functional of the electron density $\mrho$.
The average energy then reads
\[
    \Exp{\clE}
    =
    \sum_j \bar{\mphi}(\ur^{ij}) + \sum_{j} \Exp{F(\mrho(\ur^{ij},s^i),s^i)},
\]
with $\bar{\mphi}(\ur^{ij}) = \Exp{\mphi(\ur^{ij},s^i,s^j)} = \sum_{X,Y} \mphi(\ur^{ij},s^X,s^Y)$.
Now, to compute the average of the embedding term, the approximation made by \citet{smith_application_1989} that has also been used by \citet{varvenne_averageatom_2016} is to expand $F$ around the average $\bar{\mrho}$ of the electron density, and truncate the expansion after the zeroth term so that
\[
    \Exp{F(\mrho,s^i)}
    \approx
    \Exp{F(\bar{\mrho},s^i)}
    =
    \sum_X F(\bar{\mrho},s^X)
    ,
\]
with $\bar{\mrho} = \bar{\mrho}(\ur^{ij}) = \sum_X \mrho(\ur^{ij},s^X)$.

\section*{Acknowledgments}

The author benefited from stimulating discussions with Bill Curtin, Francesco Maresca, and C\'{e}line Varvenne, on applying the idea of effective media to molecular dynamics that led to the present work.
The financial support under the scope of the COMET program within the K2 Center “Integrated Computational Material, Process and Product Engineering (IC-MPPE)” (Project No 886385) is highly acknowledged. This program is supported by the Austrian Federal Ministries for Climate Action, Environment, Energy, Mobility, Innovation and Technology (BMK) and for Labour and Economy (BMAW), represented by the Austrian Research Promotion Agency (FFG), and the federal states of Styria, Upper Austria and Tyrol.

\printbibliography[heading=bibintoc]

\begin{refsection}
    \clearpage

\onecolumn

\setcounter{section}{0}
\setcounter{equation}{0}
\setcounter{figure}{0}
\setcounter{table}{0}
\setcounter{page}{1}
\setcounter{example}{0}
\setcounter{footnote}{0}

\renewcommand{\theequation}{S\arabic{equation}}
\renewcommand{\thefigure}{S\arabic{figure}}
\renewcommand{\thesection}{S\arabic{section}}
\renewcommand{\thetable}{S\arabic{table}}
\renewcommand{\thepage}{S\arabic{page}}
\renewcommand{\theexample}{S\arabic{example}}

\emptythanks

\title{Exact average many-body interatomic interaction model for random alloys \\[0.5em] Supplementary material}

\maketitle

\noindent%
In the following, we refer to any section, equation, figure, example, and algorithm, from the supplementary material as Section SX, equation (SX), Figure SX, Example SX, and Algorithm SX.
Any equation, figure, and example, from the main text is referred to as, equation (X), Figure X, and example X.

\section{Method of set partitions}
\label{sec:set_partitions}

In the following, we will re-derive the set partitioning method of \citet{mccullagh_tensor_2018} (Section 3.6) in some detail because we could not find applications of it in the atomistic modeling literature using a language that is convenient to computational scientists, and because we could not find an extension to \emph{statistical moments of tensors}.
The latter extension is conceptually straightforward, but requires some care when computing tensor products of averages because the indices of the resulting tensor may have been shuffled (see the third term on the right hand side in \eqref{eq:expansion_d=3}).
Therefore, we use the index notation in the following in order to capture this index shuffling, in comparison to the main text where we have mostly used the more concise direct notation.

\begin{figure}
    \centering
    \includegraphics[width=0.7\textwidth]{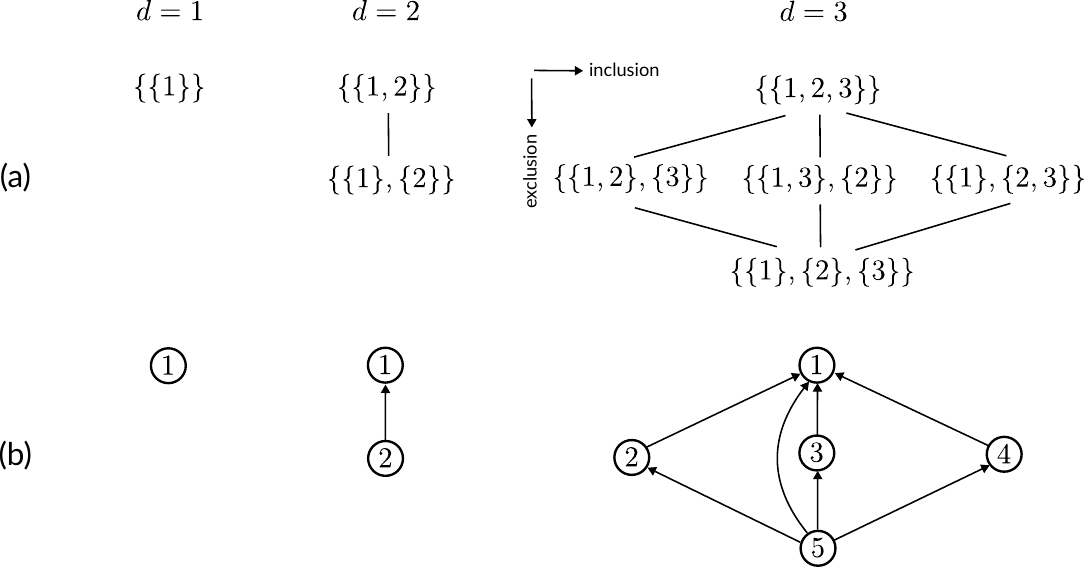}
    \caption{(a) Set partitions for $d=1,\ldots,3$. (b) Hasse diagrams corresponding to the set partitions in (a)}
    \label{fig:hasse}
\end{figure}

Our goal is therefore to find a partition of the set $\scI_d = \{ 1,\ldots,d \}$ that contains the positive integers up to the body-order of the potential, corresponding to the indices of the feature vectors.
A partition $\scP_d$ of $\scI_d$ is the set of all unique non-empty subsets of $\scI_d$ so that each element of $\scI_d$ is included exactly once in each of those subsets.
The partition $\scP_d$ can be constructed recursively using the inclusion-exclusion principle (cf., e.g., \cite{andrews_symbolic_2000}), starting from a partition $\scP_{d-1}$ by first including the next element, $d$, in all subsets of $\scP_{d-1}$, and, second, by adding it as \emph{disjoint set} to all elements of $\scP_{d-1}$ (exclusion).
This procedure is exemplified in the following for partitions up to $d = 3$ (cf., Figure \ref{fig:hasse} (a)):
\begin{align*}
    \scP_1
    &=
    \{ \{ 1 \} \},
    \\
    \scP_2
    &=
    \{ \{ 1, 2 \}, \{ \{ 1 \}, \{ 2 \} \} \},
    \\
    \scP_3
    &=
    \{ \{ 1, 2, 3 \}, \{ \{ 1, 3 \}, \{ 2 \} \}, \{ \{ 1 \}, \{ 2, 3 \} \}, \{ \{ 1, 2 \}, \{ 3 \} \}, \{ \{ 1 \}, \{ 2 \}, \{ 3 \} \} \}
    .
\end{align*}
It is easy to see that the distribution of the indices over the blocks of the elements of $\scP_2$ and $\scP_3$ correspond to how the feature vectors are assigned to the disjoint sums in \eqref{eq:expansion_d=2} and \eqref{eq:expansion_d=3}, respectively.

In order to derive a generic formula for the expansion of higher-order moments, we require some additional notation for indexing the partitions.
In what follows, each element of $p \in \scP_d$, that is, each subset of $\scI_d$, will be called a block.
Each $p$ consists of $\mu^p$ blocks, $b_1^p,\ldots,b_\mu^p$.
Further, each block $b^p \in p$ consists of $\nu^{p,b}$ indices $\lambda^{p,b} \in \scI_d$.
For the sake of clarity of the presentation, we will omit the superscripted indices $p$ and $b$ in the following if it is clear from the context to which partition and block we are referring to.
An element of $p \in \scP_d$ can then be fully indexed as follows
\[
    p
    =
    \{ \underbrace{\{ \lambda_1^1, \ldots, \lambda_\nu^1 \} }_{\textstyle b_1}, \ldots,
       \underbrace{ \{ \lambda_1^\mu, \ldots, \lambda_\nu^\mu \} }_{\textstyle b_\mu} \}
    .
\]

With the notation above, we are now able to represent an arbitrary tensor product of sums in terms of disjoint sums as follows
\begin{equation}\label{eq:disjoint_part}
    \prod_{l = 1}^d v_{k_l}^l
    =
    \prod_{l = 1}^d \sum_j v_{k_l}^{l,j}
    =
    \sum_{p \in \scP_d}
    \sum_{j_1 \neq \ldots \neq j_\mu}
    V_{k_{\lambda_1^1} \ldots k_{\lambda_\nu^1}}^{p,1,j_1}
    \ldots
    V_{k_{\lambda_1^\mu} \ldots k_{\lambda_\nu^\mu}}^{p,\mu,j_\mu}
    ,
\end{equation}
where the $V$'s are the tensor products of the feature vectors for each block $b$
\[
    V_{k_{\lambda_1^b} \ldots k_{\lambda_\nu^b}}^{p,b,j_b}
    =
    \prod_{l \in b} v_{k_l}^{l,j_b}
    .
\]
Proceeding analogously as above, we can then write the average of \eqref{eq:disjoint_part} as follows
\begin{equation}\label{eq:average_moments_disjoint_sums}
    \Exp{ \prod_{l = 1}^d v_{k_l}^l }_{Y_1,\ldots,Y_d}
    =
    \sum_{p \in \scP_d}
    \sum_{j_1 \neq \ldots \neq j_\mu}
    \Exp{ V_{k_{\lambda_1^1} \ldots k_{\lambda_\nu^1}}^{p,1,j_1} }_{Y_1}
    \ldots
    \Exp{ V_{k_{\lambda_1^\mu} \ldots k_{\lambda_\nu^\mu}}^{p,\mu,j_\mu} }_{Y_\mu}
    .
\end{equation}

The latter expression is, however, not convenient to compute because each term in the expansion scales exponentially with the number of blocks.
Fortunately, we can apply the same strategy as above in the reverse way, i.e., converting terms of disjoint sums to terms of products of sums that scale linearly with the size of the neighborhood.

A way to achieve this is to interpret a partition $\scP_d$ as a Hasse diagram, a directed acyclic graph, and deduce the arrangements of the expansion terms from the inverse of the adjacency matrix of the Hasse diagram.
The Hasse diagrams corresponding to the set partitions for $d=1,\ldots,3$ are shown in Figure \ref{fig:hasse}.
We construct the Hasse diagram from $\scP_d$ by considering the blocks as vertices.
In order to construct the edges, we define a partial order on the graph, that is, for vertices $p$ and $\widetilde p$ we say that $p \le \widetilde p$ whenever $p$ is a subpartition of $\widetilde p$;
for example, in $\scP_3$, the set $\{ \{ 1 \}, \{ 2 \}, \{ 3 \} \}$ is a subpartition of $\{ \{ 1,2 \}, \{ 3 \} \}$, but $\{ \{ 1 \}, \{ 2,3 \} \}$ is not because $\{ 2,3 \}$ is not element of $\{ \{ 1,2 \}, \{ 3 \} \}$.
If $p \le \widetilde p$, then there exists an edge that points from $p$ to $\widetilde p$ (cf., Figure \ref{fig:hasse}).

Now, to illustrate the procedure of converting between products of sums and disjoint sums using Hasse diagrams, we define two functions on $\scP_d$, $f$, and $g$, where $f$ is a function of products sums, and $g$ is a function of disjoint sums, such that
\begin{align*}
    f(\widetilde p)
    =
    \prod_{\widetilde b \in \widetilde p}
    \sum_{j_{\widetilde b}}
    \widetilde\chi^{\widetilde p,\widetilde b,j_{\widetilde b}}
    ,
    &&
    g(p)
    =
    \sum_{j_1 \neq \ldots \neq j_\mu}
    \chi^{p,1,j_1}
    \ldots
    \chi^{p,\mu,j_\mu}
    ,
\end{align*}
with the $\widetilde\chi$'s being generic functions, and the $\chi$'s being defined through the relation
\begin{equation}\label{eq:partition_relation}
    f(\widetilde p) = \sum_{\substack{p \in \scP,\\ p \le \widetilde p}} g(p)
    ,
\end{equation}
which is nothing but a (non-tensorial) generalization of \eqref{eq:disjoint_part}.
We can write \eqref{eq:partition_relation} in matrix-vector notation as
\[
    \uf = \uuA \ug
    ,
\]
where $\uuA$ is the adjacency matrix, and
\begin{align*}
    \uf =
    \begin{pmatrix} f(\widetilde p_1) & \cdots & f(\widetilde p_n) \end{pmatrix}^\sT
    ,
    &&
    \ug =
    \begin{pmatrix} g(p_1) & \cdots & g(p_n) \end{pmatrix}^\sT
    ,
\end{align*}
with the $p$'s being ordered such that $p_1 \le \ldots \le p_n$.
According to \eqref{eq:partition_relation}, the adjacency matrix is then a lower triangular matrix.
Moreover, $\uuA$ is regular, so we can convert a disjoint sum back to a product of sums via the inverse relation
\[
    \ug = \uuA^{-1} \uf
    .
\]
This implies that we can expand some function of disjoint sums $g$ as follows
\[
    \sum_{j_1 \neq \ldots \neq j_\mu}
    \chi^{p,1,j_1}
    \ldots
    \chi^{p,\mu,j_\mu}
    =
    \sum_{\substack{\widetilde p \in \scP\\ \widetilde p \le p}} A^{-1}_{p\widetilde p}
    \prod_{\widetilde b \in \widetilde p}
    \sum_{j_{\widetilde b}}
    \widetilde\chi^{\widetilde p,\widetilde b,j_{\widetilde b}}
    ,
\]
where $A^{-1}_{p\widetilde p}$ are the elements of the inverse of $\uuA$,%
\footnote{In the statistics literature, the matrices $\uuA$ and $\uuA^{-1}$ are usually referred to as Zeta matrix and M\"{o}bius matrix, respectively}
and
\[
    \widetilde\chi^{\widetilde p,\widetilde b,j}
    =
    \prod_{b \in \widetilde b}
    \chi^{p,b,j}
    .
\]

Since $\uuA$ is constant, we have $\Exp{\ug} = \uuA^{-1} \Exp{\uf}$.
Now, to represent a disjoint sum of averages as products of averages, we simply interpret the application of the averaging operator on $f$ and $g$ as new functions $f'$ and $g'$, i.e.,
\[
    \Exp{\ug} = \ug' = \uuA^{-1} \Exp{\uf} = \uuA^{-1} \uf'
    .
\]
Using this observation, we are now able to convert the disjoint sums in \eqref{eq:average_moments_disjoint_sums} into products of sums leading to
\[
    \sum_{j_1 \neq \ldots \neq j_\mu}
    \Exp{V_{k_{\lambda_1^1} \ldots k_{\lambda_\nu^1}}^{p,1,j_1}}_Y
    \ldots
    \Exp{V_{k_{\lambda_1^\mu} \ldots k_{\lambda_\nu^\mu}}^{p,\mu,j_\mu}}_Y
    =
    \sum_{\substack{\widetilde p \in \scP_d\\ \widetilde p \le p}}
    A^{-1}_{p\widetilde p}
    \prod_{\widetilde b \in \widetilde p}
    \sum_{j_{\widetilde b}}
    \Exp{
    \widetilde V^{\widetilde p,\widetilde b,j_{\widetilde b}}_{k_{\widetilde\lambda_\nu} \ldots k_{\widetilde\lambda_\nu}}
    }_Y
    ,
\]
with
\[
    \Exp{
    \widetilde V^{\widetilde p,\widetilde b,j}_{k_{\widetilde\lambda_\nu} \ldots k_{\widetilde\lambda_\nu}}
    }_Y
    =
    \prod_{b \in \widetilde b}
    \Exp{
    V_{k_{\lambda_1} \ldots k_{\lambda_\nu}}^{p,b,j}
    }_Y
    .
\]

We are now in a position to write the energy averaged over all neighboring species in terms of products of sums as follows
\[
    \Exp{\clE}_Y
    =
    T_{k_1, \ldots, k_d}
    \sum_{p \in \scP_d}
    \sum_{\substack{\widetilde p \in \scP_d\\ \widetilde p \le p}}
    L^{-1}_{p\widetilde p}
    \prod_{\widetilde b \in \widetilde p}
    \sum_{j_{\widetilde b}}
    \Exp{
    \widetilde V^{\widetilde p,\widetilde b,j_{\widetilde b}}_{k_{\widetilde\lambda_1} \ldots k_{\widetilde\lambda_\nu}}
    }_Y
    .
\]
Then, the only piece that is left is taking the average also over all central species.
This yields the exact formula for the average energy of multilinear interatomic interaction models \eqref{eq:multilinear_potential}
\begin{equation}\label{eq:average_multilinear_pot_ext}
\boxed{
    \begin{aligned}
        \Exp{\clE}
        &=
        T_{k_1, \ldots, k_d}
        \sum_{p \in \scP_d}
        \sum_{\substack{\widetilde p \in \scP_d\\ \widetilde p \le p}}
        A^{-1}_{p\widetilde p}
        \Exp{
        \prod_{\widetilde b \in \widetilde p}
        \sum_{j_{\widetilde b}}
        \Exp{\widetilde V^{\widetilde p,\widetilde b,j_{\widetilde b}}_{k_{\widetilde\lambda 1} \ldots k_{\widetilde\lambda_\nu}}}_Y
        }_X
        \\
        &=
        T_{k_1, \ldots, k_d}
        \sum_{p \in \scP_d}
        \sum_{\substack{\widetilde p \in \scP_d\\ \widetilde p \le p}}
        A^{-1}_{p\widetilde p}
        \Exp{
        \prod_{\widetilde b \in \widetilde p}
        \sum_{j_{\widetilde b}}
        \prod_{b \in \widetilde b}
        \Exp{V_{k_{\lambda_1^b} \ldots k_{\lambda_\nu^b}}^{p,b,j_{\widetilde b}}}_Y
        }_X
        \\
        &=
        T_{k_1, \ldots, k_d}
        \sum_{p \in \scP_d}
        \sum_{\substack{\widetilde p \in \scP_d\\ \widetilde p \le p}}
        A^{-1}_{p\widetilde p}
        \Exp{
        \prod_{\widetilde b \in \widetilde p}
        \sum_{j_{\widetilde b}}
        \prod_{b \in \widetilde b}
        \Exp{\prod_{l \in b} v_{k_l}^{l,j_{\widetilde b}}}_Y
        }_X
        .
    \end{aligned}
}
\end{equation}
For a given body-order and a given number of species, the model \eqref{eq:average_multilinear_pot} scales linearly with the neighborhood size.

\begin{example}
    Specifically, for feature vectors of type \eqref{eq:example_feature_vector}, the average energy is given as follows
    \[
        \Exp{\clE}
        =
        T_{k_1, \ldots, k_d}
        \sum_{p \in \scP_d}
        \sum_{\widetilde p \in \scP_d}
        A^{-1}_{p\widetilde p}
        \sum_X c_X
        \prod_{\widetilde b \in \widetilde p}
        \sum_{j_{\widetilde b}}
        \prod_{b \in \widetilde b}
        \sum_Y c_Y \prod_{l \in b} Q_{\alpha_l}(\vert \ur^{ij} \vert) z_{\beta_l}^X z_{\gamma_l}^Y Y_{\ell_l m_l}(\widehat\ur^{ij})
        .
    \]
\end{example}

We would further like to point out that the construction of the expansion can be automated, as shown by Algorithm \ref{algo:hasse}, which only requires a few operations from elementary set and graph theory.

\begin{algorithm}[hbt]
    \SetAlgoSkip{bigskip}
    \LinesNumbered
    \SetKwInput{Input}{Input}
    \SetKwInput{Output}{Output}
    \caption{Construct set partition and Hasse diagram for an arbitrary body-order (\textbf{Hasse})}
    \label{algo:hasse}
    \Input{Tensor order $d$}
    $\scP_d \,\leftarrow\, \{ \{ 1 \} \}$;\\
    \For{$i = 2, \ldots, d$}{
        $\scP_d' \,\leftarrow\, \emptyset$;\\
        \ForEach{$p \in \scP_d$}{
            \ForEach{$b \in p$}{
                $b' \,\leftarrow\, b \cup i$;\\
                $p' \,\leftarrow\, (p \setminus b) \cup b'$;\\
                $\scP_d' \,\leftarrow\, \scP_d' \cup p'$; \tcp{inclusion}
            }
            $\scP_d' \,\leftarrow\, \scP_d' \cup (p \cup i)$; \tcp{exclusion}
        }
        $\scP_d \,\leftarrow\, \scP_d'$;\\
    }
    $G \,\leftarrow\, \text{construct Hasse diagram from $\scP_d$}$;\\
    \Output{partition $\scP_d$, adjacency matrix $\uuA$ of $G$}
\end{algorithm}

\section{Linear-scaling algorithm for computing average energies}
\label{sec:algo}

We now describe the algorithm for computing average per-atom energies that scales linearly with the size of an atomic neighborhood using ETN potentials.
In the proposed algorithm, there are three levels of tensorial contractions for each expansion term in \eqref{eq:average_multilinear_pot} (i.e., for each partition $\widetilde p$):
At the lowest level, level 1, we perform per-neighborhood-species operations, one level higher, at level 2, we perform per-neighborhood-atom operations, and at the highest level, level 3, we perform per-central-atom operations.
Algorithm \ref{algo:ave-etn} shows the main steps of computing the average energy of an ETN potential in symbolic notation.
In the following, we describe those steps more conveniently using our tensor diagram notation as outlined in the Methods Section from the main text.

\begin{algorithm}[hbt]
    \SetAlgoSkip{bigskip}
    \LinesNumbered
    \SetKwInput{Input}{Input}
    \SetKwInput{Output}{Output}
    \caption{Compute average energy of an ETN potential of order $d$}
    \label{algo:ave-etn}
    \Input{Neighborhood $\{ \ur^{ij} \}$, tensor order $d$}
    $(\scP_d, \uuA) \,\leftarrow\, \textbf{Hasse}(d)$;\\
    \ForEach{$X$}{
        \ForEach{$p \in \scP_d, \; \widetilde p \in \{\, \widetilde p^{\,\prime} \in \scP_d \,\vert\, \widetilde p^{\,\prime} \le p \,\}$}{
            \ForEach{$\widetilde b \in \widetilde p$}{
                \ForEach{$\ur^{ij} \in \{ \ur^{ij} \}$}{
                    $u_{\ell m \alpha}^j \,\leftarrow\, Y_{\ell m}(\widehat{\ur}^{ij}) Q_\alpha{(\vert \ur^{ij} \vert)}$; \tcp{compute angular \& radial basis}
                    \ForEach{$Y$}{
                        $v_{(\ell m n)}^j \,\leftarrow\, B_{\ell n\alpha\eta} u_{\ell m \alpha}^j A_{\ell\eta\beta\gamma} z_\beta^X z_\gamma^Y$;\\
                        \ForEach{$b \in \widetilde b$}{
                            \tcp{Level 1: per-neighborhood-species operations}
                            $
                                S_{r_{l_1} \ldots r_{l_{\nu + 1}}}^{b,j}
                                \leftarrow
                                S_{r_{l_1} \ldots r_{l_{\nu + 1}}}^{b,j}
                                +
                                c_Y
                                T_{r_{l_1} k_{l_1} r_{l_2}}^{l_1} \ldots T_{r_{l_\nu} k_{l_\nu} r_{l_\nu + 1}}^{l_\nu} v_{k_{l_1}}^{l_1,j} \ldots v_{k_{l_\nu}}^{l_\nu,j}
                            $
                        }
                    }
                    \tcp{Level 2: per-neighborhood-atom operations}
                    $
                        S_{r_{l_1} \ldots r_{l_{\nu + 1}}}^{\widetilde b}
                        \leftarrow
                        S_{r_{l_1} \ldots r_{l_{\nu + 1}}}^{\widetilde b}
                        +
                        S_{r_{l_1} \ldots r_{l_{\nu + 1}}}^{b_1,j}
                        \ldots
                        S_{r_{l_1} \ldots r_{l_{\nu + 1}}}^{b_\mu,j}
                    $
                }
            }
            \tcp{Level 3: per-central-atom operations}
            $
                \clE
                \leftarrow
                \clE
                +
                c_X
                A^{-1}_{p\widetilde p}
                \left(
                S_{r_{l_1} \ldots r_{l_{\nu + 1}}}^{\widetilde b_1}
                \ldots
                S_{r_{l_1} \ldots r_{l_{\nu + 1}}}^{\widetilde b_\mu}
                \right)
            $
            ; \tcp{scalar contraction}
        }   
    }
    \Output{$\clE$}
\end{algorithm}

At the first level, we contract the coefficient tensors with the per-neighborhood feature vectors, i.e., for each neighborhood atom $j$ and species $Y$.
This contraction yields a tensor
\[
    S_{\langle \bullet \rangle}^{b,j}
    =
    \begin{aligned}
        \includegraphics[scale=0.75]{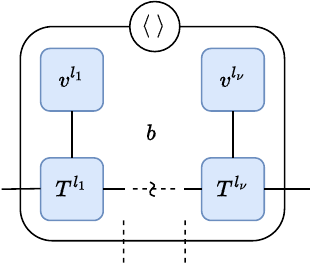}
    \end{aligned}
\]
for each block $b \in \widetilde b$, where $\langle\,\rangle$ denotes the average operator that acts on the encircled tensor.
Note that, when $d>2$, there will be blocks for which the indices are not consecutive, and those blocks yield higher-order tensors of sizes depending on the ranks;
this is visualized in the diagram above using dashed links leaving the average operator.

At the second level, we contract all tensors $S_{\langle \bullet \rangle}^{b,j}$ that we have computed at the first level.
That is, for each neighboring atom $j$ we compute
\[
    S_{\langle \bullet \rangle}^{\widetilde b,j}
    =
    \begin{aligned}
        \includegraphics[scale=0.75]{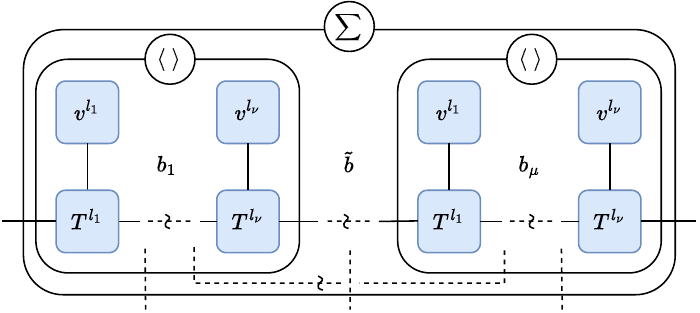}
    \end{aligned}
\]
and assemble them to $S_{\langle \bullet \rangle}^{\widetilde b} = \sum_j S_{\langle \bullet \rangle}^{\widetilde b,j}$, visualized using the $\sum$ symbol.

At the third level, we contract all tensors $S_{\langle \bullet \rangle}^{\widetilde b}$ corresponding to blocks $\widetilde b$ that belong to a given expansion term $\widetilde p$.
This operation can be represented in diagrammatic notation as follows
\[
    \begin{aligned}
        \includegraphics[scale=0.75]{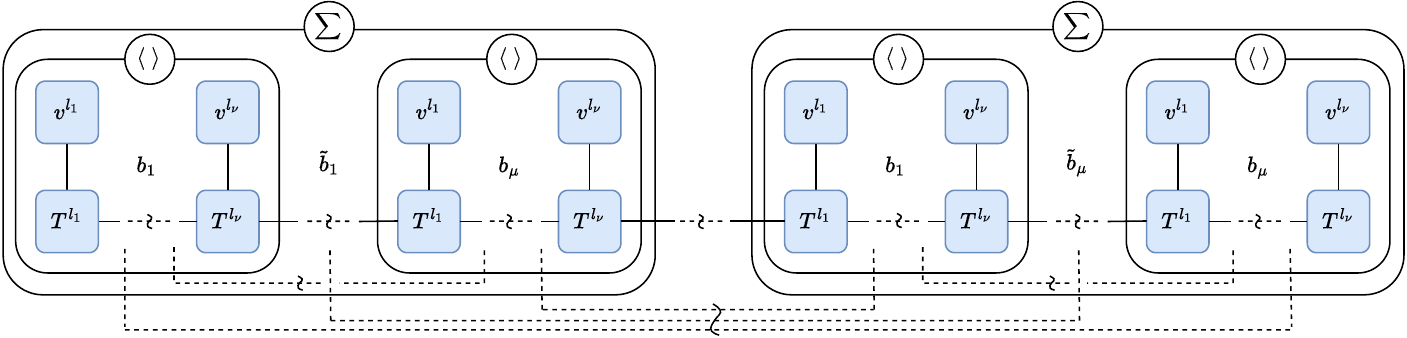}
    \end{aligned}
    .
\]
The results of these contractions will be scalars that add up to the average per-atom energy.
The previous steps are then repeated for all $X$ in order to compute the average over all possible species for the central atom $i$.

While the algorithm presented here computes the per-atom energy $\clE$, it is straightforward to additionally compute average forces and stresses using automatic differentiation;
we therefore omit the technical details for the sake of compactness.

At this point, we remark that, if constructing those higher-order tensors becomes a bottleneck, one may transpose the indices of $T_{\langle d \rangle}$ to align all $T$'s involved in the contraction in a preprocessing step.
In the main text, we have assumed that it can be done and simply applied braiding operators to $T_{\langle d \rangle}$ (cf., eq. \eqref{eq:average_multilinear_pot}).
This, however, requires re-factorizing the involved tensors, but can be done efficiently using existing algorithms for conventional tensor networks (cf. \cite{cichocki_tensor_2017,hodapp_equivariant_2023});
nonetheless, these algorithms still need to be adapted and implemented for ETNs.
Assuming that the indices to be contracted are in consecutive order, the contraction of the feature vectors at level 1 then always yields a $r_{l_1} \times r_{l_\nu + 1}$ matrix $\uuS^{b,j}$, and likewise at level 2 (cf., Figure \ref{fig:average_etn_aligned}).

\section{Average four-body potentials}
\label{sec:four_body}

In the following, we will elucidate on how to apply the procedure above by deriving functional forms of the exact averages of multilinear potentials with up-to four-body interactions.
For four-body potentials, the adjacency matrix and its inverse are given by
\begin{align*}
    \uuA
    =
    \begin{pmatrix}
        1 &&&& \\ 1 & 1 &&& \\ 1 && 1 && \\ 1 &&& 1 & \\ 1 & 1 & 1 & 1 & 1
    \end{pmatrix}
    ,
    &&
    \uuA^{-1}
    =
    \begin{pmatrix}
        1 &&&& \\ -1 & 1 &&& \\ -1 && 1 && \\ -1 &&& 1 & \\ 2 & -1 & -1 & -1 & 1
    \end{pmatrix}
    .
\end{align*}
In \eqref{eq:expansion_d=3}, there are four terms with disjoint sums that we have convert into terms of products of sums using $\uuA^{-1}$.
For the first three terms, we obtain
\begin{align*}
    \sum_{j_1} \Exp{v^{1,j_1}_{k_1} v^{2,j_1}_{k_2}}_Y \sum_{j_3 \neq j_1} \Exp{v^{3,j_3}_{k_3}}_Y
    =
    g'_{k_1 k_2 k_3}(p_2)
    &=
    - f'_{k_1 k_2 k_3}(p_1) + f'_{k_1 k_2 k_3}(p_2)
    \\
    &=
    - \sum_{j_1} \Exp{v^{1,j_1}_{k_1} v^{2,j_1}_{k_2}}_Y \Exp{v^{3,j_1}_{k_3}}_Y
    + \sum_{j_1} \Exp{v^{1,j_1}_{k_1} v^{2,j_1}_{k_2}}_Y \sum_{j_3} \Exp{v^{3,j_3}_{k_3}}_Y
    ,
    \\
    \sum_{j_1} \Exp{v^{1,j_1}_{k_1} v^{2,j_1}_{k_2}}_Y \sum_{j_3 \neq j_1} \Exp{v^{3,j_3}_{k_3}}_Y
    =
    g'_{k_1 k_2 k_3}(p_3)
    &=
    - f'_{k_1 k_2 k_3}(p_1) + f'_{k_1 k_2 k_3}(p_3)
    \\
    &=
    - \sum_{j_1} \Exp{v^{1,j_1}_{k_1}}_Y \Exp{v^{2,j_2}_{k_2} v^{3,j_2}_{k_3}}_Y
    + \sum_{j_1} \Exp{v^{1,j_1}_{k_1}}_Y \sum_{j_2} \Exp{v^{2,j_2}_{k_2} v^{3,j_2}_{k_3}}_Y
    ,
    \\
    \sum_{j_1} \Exp{v^{1,j_1}_{k_1}}_Y \sum_{j_2 \neq j_1} \Exp{v^{2,j_2}_{k_2} v^{3,j_2}_{k_3}}_Y
    =
    g'_{k_1 k_2 k_3}(p_4)
    &=
    - f'_{k_1 k_2 k_3}(p_1) + f'_{k_1 k_2 k_3}(p_4)
    \\
    &=
    - \sum_{j_1} \Exp{v^{1,j_1}_{k_1} v^{3,j_1}_{k_3}}_Y  \Exp{v^{2,j_1}_{k_2}}_Y
    + \sum_{j_1} \Exp{v^{1,j_1}_{k_1} v^{3,j_1}_{k_3}}_Y  \sum_{j_2} \Exp{v^{2,j_2}_{k_2}}_Y
    ,
\end{align*}
and for the last term containing three disjoint sums
\begin{multline*}
    \sum_{j_1} \Exp{v^{1,j_1}_{k_1}}_Y \sum_{j_2 \neq j_1} \Exp{v^{2,j_2}_{k_2}}_Y \sum_{j_3 \neq j_2} \Exp{v^{3,j_3}_{k_3}}_Y
    =
    g'_{k_1 k_2 k_3}(p_5)
    \\
    \begin{aligned}
    &=
    2f'_{k_1 k_2 k_3}(p_1) - f'_{k_1 k_2 k_3}(p_2) - f'_{k_1 k_2 k_3}(p_3) - f'_{k_1 k_2 k_3}(p_4) + f'_{k_1 k_2 k_3}(p_5)
    \\
    &=
      2 \sum_{j_1} \Exp{v^{1,j_1}_{k_1}}_Y \Exp{v^{2,j_1}_{k_2}}_Y  \Exp{v^{3,j_1}_{k_3}}_Y
    -   \sum_{j_1} \Exp{v^{1,j_1}_{k_1}}_Y \Exp{v^{2,j_1}_{k_2}}_Y \sum_{j_3} \Exp{v^{3,j_3}_{k_3}}_Y
    -   \sum_{j_1} \Exp{v^{1,j_1}_{k_1}}_Y \sum_{j_2} \Exp{v^{2,j_1}_{k_2}}_Y \Exp{v^{3,j_2}_{k_3}}_Y
    \end{aligned}
    \\
    -   \sum_{j_1} \Exp{v^{1,j_1}_{k_1}}_Y \Exp{v^{3,j_1}_{k_3}}_Y \sum_{j_2} \Exp{v^{2,j_2}_{k_2}}_Y
    +   \sum_{j_1} \Exp{v^{1,j_1}_{k_1}}_Y \sum_{j_2} \Exp{v^{2,j_2}_{k_2}}_Y \sum_{j_3} \Exp{v^{3,j_3}_{k_3}}_Y
    .
\end{multline*}
The full expression of the average energy then reads
\begin{multline*}
    \Exp{\clE}
    =
    T_{k_1 k_2 k_3}
    \Bigg\<
        \sum_j \Exp{v^{1,j}_{k_1} v^{2,j}_{k_2} v^{3,j}_{k_3}}_Y
    -   \sum_{j_1} \Exp{v^{1,j_1}_{k_1} v^{2,j_1}_{k_2}}_Y \Exp{v^{3,j_1}_{k_3}}_Y
    +   \sum_{j_1} \Exp{v^{1,j_1}_{k_1} v^{2,j_1}_{k_2}}_Y \sum_{j_3} \Exp{v^{3,j_3}_{k_3}}_Y \\
    -   \sum_{j_1} \Exp{v^{1,j_1}_{k_1}}_Y \Exp{v^{2,j_2}_{k_2} v^{3,j_2}_{k_3}}_Y
    +   \sum_{j_1} \Exp{v^{1,j_1}_{k_1}}_Y \sum_{j_2} \Exp{v^{2,j_2}_{k_2} v^{3,j_2}_{k_3}}_Y
    -   \sum_{j_1} \Exp{v^{1,j_1}_{k_1} v^{3,j_1}_{k_3}}_Y \Exp{v^{2,j_1}_{k_2}}_Y \\
    +   \sum_{j_1} \Exp{v^{1,j_1}_{k_1} v^{3,j_1}_{k_3}}_Y \sum_{j_2} \Exp{v^{2,j_2}_{k_2}}_Y
    + 2 \sum_{j_1} \Exp{v^{1,j_1}_{k_1}}_Y \Exp{v^{2,j_1}_{k_2}}_Y  \Exp{v^{3,j_1}_{k_3}}_Y
    -   \sum_{j_1} \Exp{v^{1,j_1}_{k_1}}_Y \Exp{v^{2,j_1}_{k_2}}_Y \sum_{j_3} \Exp{v^{3,j_3}_{k_3}}_Y \\
    -   \sum_{j_1} \Exp{v^{1,j_1}_{k_1}}_Y \sum_{j_2} \Exp{v^{2,j_1}_{k_2}}_Y \Exp{v^{3,j_2}_{k_3}}_Y
    -   \sum_{j_1} \Exp{v^{1,j_1}_{k_1}}_Y \Exp{v^{3,j_1}_{k_3}}_Y \sum_{j_2} \Exp{v^{2,j_2}_{k_2}}_Y \\
    +   \sum_{j_1} \Exp{v^{1,j_1}_{k_1}}_Y \sum_{j_2} \Exp{v^{2,j_2}_{k_2}}_Y \sum_{j_3} \Exp{v^{3,j_3}_{k_3}}_Y
    \Bigg\>_X
    .
\end{multline*}

\printbibliography[heading=subbibliography]
\end{refsection}

\end{document}